\documentclass[journal]{IEEEtran}

\usepackage[colorlinks,urlcolor=blue,linkcolor=blue,citecolor=blue]{hyperref}
\usepackage{cite}
\usepackage{amsmath,amssymb,amsfonts}
\usepackage{graphicx}
\usepackage{textcomp}
\usepackage{xcolor}
\usepackage{amsmath}
\usepackage{float} 
\usepackage{tikz}
\usetikzlibrary{%
    shapes,%
    calc%
}

\usepackage{booktabs}
\usepackage{makecell}

\usepackage[ruled,linesnumbered]{algorithm2e}
 \usepackage{algpseudocode}
\usepackage[normalem]{ulem}
\usepackage{multirow}
\usepackage{subcaption}

\def\BibTeX{{\rm B\kern-.05em{\sc i\kern-.025em b}\kern-.08em
    T\kern-.1667em\lower.7ex\hbox{E}\kern-.125emX}}
    
\makeatletter
\let\original@algocf@latexcaption\algocf@latexcaption
\long\def\algocf@latexcaption#1[#2]{%
	\@ifundefined{NR@gettitle}{%
		\def\@currentlabelname{#2}%
	}{%
		\NR@gettitle{#2}%
	}%
	\original@algocf@latexcaption{#1}[{#2}]%
}
\makeatother
   
\begin{document}

\title{A Deep Q-Learning based, Base-Station Connectivity-Aware, Decentralized Pheromone Mobility Model for Autonomous UAV Networks}

\author{Shreyas~Devaraju\thanks{S. Devaraju is with Computational Science, San Diego State University \& University of California, Irvine, CA, USA, sdevaraju@sdsu.edu},~\IEEEmembership{Student~Member,~IEEE},
\and Alexander~Ihler\thanks{A. Ihler is with School of Information \& Computer Science, University of California, Irvine, CA, USA, ihler@ics.uci.edu},~\IEEEmembership{Member,~IEEE},
\and Sunil~Kumar\thanks{S. Kumar is with Electrical \& Computer Engineering Department, San Diego State University, San Diego, CA USA, skumar@sdsu.edu},~\IEEEmembership{Senior~Member,~IEEE} 
}

\maketitle

\begin{abstract}
UAV networks consisting of low SWaP (size, weight, and power), fixed-wing UAVs are used in many applications, including area monitoring, search and rescue, surveillance, and tracking. Performing these operations efficiently requires a scalable, decentralized, autonomous UAV network architecture with high network connectivity.  Whereas fast area coverage is needed for quickly sensing the area, strong node degree and base station (BS) connectivity are needed for UAV control and coordination and for transmitting sensed information to the BS in real time. However, the area coverage and connectivity exhibit a fundamental trade-off: maintaining connectivity restricts the UAVs' ability to explore. 

In this paper, we first present a node degree and BS connectivity-aware distributed pheromone (BS-CAP) mobility model to autonomously coordinate the UAV movements in a decentralized UAV network. This model maintains a desired connectivity among 1-hop neighbors and to the BS while achieving fast area coverage. Next, we propose a deep Q-learning policy based BS-CAP model (BSCAP-DQN) to further tune and improve the coverage and connectivity trade-off. 
Since it is not practical to know the complete topology of such a network in real time, the proposed mobility models work online, are fully distributed, and rely on neighborhood information. 
Our simulations demonstrate that both proposed models achieve efficient area coverage and desired node degree and BS connectivity,
improving significantly over existing schemes.
\end{abstract}


\section{Introduction}
Unmanned aerial vehicles (UAVs) equipped with self-localization and sensing capabilities have become popular for applications such as search-and-rescue, surveillance and area monitoring, and target tracking \cite{SAR,bsCon,b2}.
We focus on low SWaP (size, weight, and power), fixed-wing UAVs, which offer a balance of portability and area coverage (with higher speeds and longer lifetimes than rotor-based UAVs).
However, low SWaP UAVs face several practical difficulties: they have a limited range of communication and are more prone to failure than larger UAVs.  These issues motivate a \textbf{decentralized} and \textbf{autonomous} network of UAVs, in which the nodes explore an area, perform local sensing, and communicate with their neighbors without any global knowledge of network topology.  While decentralized networks are easily scalable and have no single point of failure, the autonomous node movement generates uncertain trajectories, leading to disconnected networks and unstable communication routes.
To gather the sensed information and allow real-time coordination without dedicated communications infrastructure (which is typically unavailable in the target environments), the UAVs must act so as to maintain connectivity.
In particular, they must trade off between area coverage and network connectivity, since dispersing the UAVs to improve coverage can negatively impact their connectivity, and vice-versa \cite{bsCon}. 

\begin{figure*}[h]
\centering
    \includegraphics[width=0.9\textwidth]{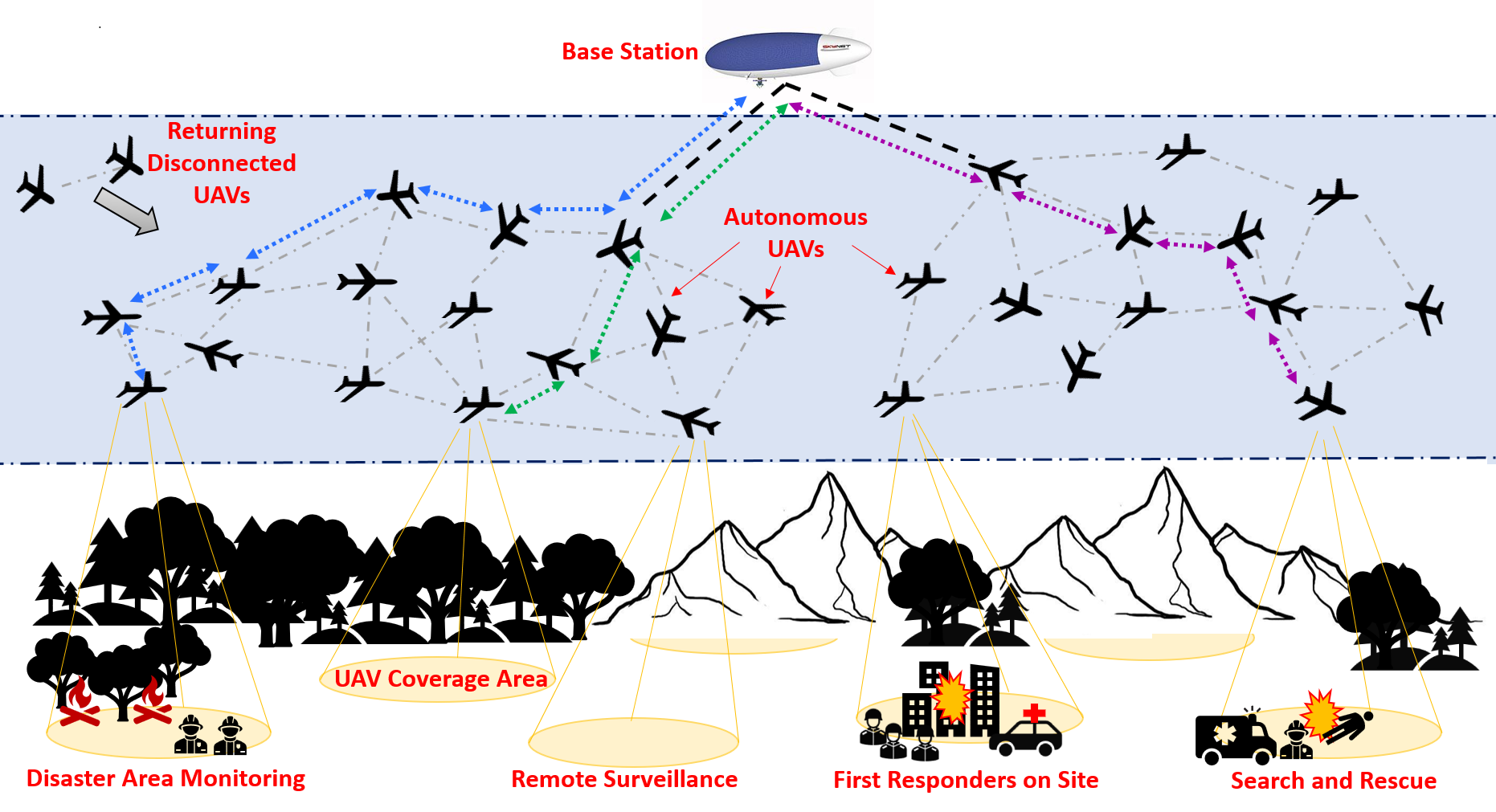}
    \caption{Decentralized, autonomous UAV network performing monitoring, search and surveillance in inaccessible disaster areas.}
    \label{fig_sar_applications}
\end{figure*}

More concretely, we consider a scenario consisting of a swarm of 30-50 low SWaP fixed-wing UAVs (e.g., \cite{locust, switchblade, lancet, coyote}) performing area monitoring and surveillance in a 6 km $\times$ 6 km area, in which no fixed communication infrastructure, such as a cellular network, is available.  
The inexpensive, low SWaP UAVs have a communication range of around 1000 m (e.g., \cite{MCA-OLSR, bsConCov, ieee80211ah}) and fly at low altitudes at speeds varying from 20 m/s to 40 m/s (e.g., \cite{switchblade}).
Fig. \ref{fig_sar_applications} illustrates an area monitoring application in an inaccessible, disaster-hit area.
With no central authority in charge of path planning, each UAV autonomously determines its trajectory based on its local neighborhood information.
The individual UAVs collaboratively explore the area to collect information, which must then be reliably sent to an aerial base station (BS). The aerial BS can then forward the information to a control center to make informed decisions in real time. 

These settings require a moderate to high density of UAVs in the target area, since too few low SWaP UAVs flying at low altitudes cannot cover the area in a limited operational time.  The network's decentralized nature enables it to be robust to loss of UAVs (due to limited battery power, hostile environments, or other failures); by acting autonomously, the network can react to such changes, including enduring periods of disconnection, and UAVs do not need to maintain communications with a pilot.
The UAVs continuously monitor the area, find objects of interest, and then report back. When they find a target of interest, they initiate a connection to the BS (located at the boundary of the monitored area) and communicate the sensed information to the BS in real time.

One common method to coordinate and control the UAV movement is to use stigmergic digital pheromones
\cite{b2,rajan-b17, c21}. Pheromone-based mobility models are simple, scalable, and robust, and can be implemented in both centralized or decentralized systems.  However, these models focus only on the area coverage while ignoring the connectivity among the UAVs. 

Many existing schemes for maintaining connectivity in UAV networks are either centralized or use hierarchical architecture or relay nodes \cite{bs24, c21, PSO_UAVwsn }. These approaches are plagued by scalability issues and delays, and are vulnerable to attacks (such as in battle-field deployments) or node failures \cite{heir_net_challenges2}.
Other methods (e.g., \cite{ b2, b16, sensor13}) modify the UAV mobility model 
to 
maintain local connectivity among the nodes.
However, few UAV mobility schemes 
attempt to address both local connectivity (node degree) and BS connectivity in a 
decentralized, non-hierarchical UAV network \cite{bsConCov, bs22 }.

In previous work, we proposed a distributed, connectivity-aware pheromone mobility model (CAP) \cite{CAP_ref} and a deep Q-learning based CAP model (CAP-DQN) \cite{CAPDQN_ref} that balance efficient area coverage with a desired node degree in autonomous UAV networks. Importantly however, it did not consider BS connectivity. 
In this work, we extend our approach to address the problem of achieving fast area coverage while maintaining both local (node degree) and BS connectivity in an autonomous, decentralized, non-hierarchical UAV network. 

We make several specific contributions beyond \cite{CAP_ref,CAPDQN_ref}:

\begin{itemize}
    \item We design a heuristic-based node degree- and BS-connectivity aware pheromone mobility model (BS-CAP), by modifying the CAP scheme \cite{CAP_ref} trajectory selection and its hello information exchanges to support BS connectivity information.  
    
    \item We then adapt and improve on our heuristic policy using reinforcement learning. 
    We use deep Q-learning to train a neural network model mapping the UAV's locally available information to each action's expected utility, and use this to select the optimal policy. Compared to the CAP-DQN scheme \cite{CAPDQN_ref}, our model includes additional state information and an updated reward function that accounts for both connectivity criteria. 
    
    \item We include two new performance measures (\emph{Percentage of Time Connected to BS} and \emph{Giant Component}), compare our method to an additional reference scheme, and evaluate the impact of node failures. 

    \item Since it is not practical to know the complete topology of a decentralized network in real-time, both mobility models work online and are fully distributed, enabling the UAVs to make autonomous decisions in an unknown environment.

    \item Both proposed models achieve a fast area coverage, strong node degree, high BS connectivity, and robustness to node failures. This provides fast information sensing from the monitored area and delivers the information to the BS in real time. 
\end{itemize}

Paper Organization: We first review several existing UAV mobility models in Section~\ref{Related_Work_section}, 
followed by an overview of the pheromone-based mobility model on which our approach is based in Section~\ref{Overview_Pheromone_Mobility_section}. We then describe our proposed node degree and BS connectivity aware pheromone (BS-CAP) mobility model in Section~\ref{proposed_BSCAP}, followed by our proposed deep Q-learning based BSCAP-DQN model in Section~\ref{DQNsection}. We evaluate the performance of our approaches in simulation and discuss the results in Section~\ref{SimultionResults}, followed by concluding remarks in Section~\ref{conclusion_section}.

\section{Related Work}\label{Related_Work_section}
Several algorithms, such as particle swarm optimization, artificial bee colony, and ant colony optimization, 
have been proposed to coordinate swarms for various search, rescue, and tracking applications \cite{pigeon_search}. One widely used approach is the use of stigmergic digital pheromones \cite{b2, rajan-b17, c21}, which act as spatio-temporal potential fields to coordinate and control the UAV movements. In digital pheromone schemes, information about the pheromone map is communicated between agents in the network through direct or indirect communication. In decentralized UAV networks, each UAV maintains a pheromone map of its neighborhood through direct communication, wherein the pheromone deposits are communicated only locally.
Some proposed schemes use a fusion of digital pheromone algorithm and flocking behaviors to coordinate a group of UAVs for performing distributed target search and tracking \cite{b2, b16}. These schemes do not maintain BS connectivity.

Decentralized and non-hierarchical UAV networks that modify the mobility model to prioritize BS connectivity include \cite{bsConCov, bs22}.
The ConCov model \cite{bsConCov} is a distributed, hybrid mobility scheme that focuses on fast area coverage and BS connectivity. It uses artificial potential fields to minimize overlap in the areas sensed by neighboring UAVs. It also uses local neighbor locations and routing information to tune its performance between area coverage and BS connectivity.  

Messous et al.~\cite{bs22} studied BS connectivity in UAV swarms by using a fuzzy inference system to compute a weight for the UAV's connectivity to its neighbors, hop count to the BS, and energy level. However, designing the fuzzy inference system can be difficult for complex systems where the relationships among different inputs are not well understood. 
The schemes proposed in \cite{sensor13,sensor14} consider coverage of ground users and reliable network connectivity. However, they do not explicitly consider BS connectivity; instead, they assume that one of the UAVs is connected to the BS (a sink or gateway node).
Another related body of literature is connectivity preservation schemes for multi-robot systems using algebraic connectivity \cite{survey_multi-robots1, survey_multi-robots3}. These schemes assume the network is initially connected and do not typically consider BS connectivity.

Another set of approaches uses a centralized or hierarchical network architecture. 
A centralized scheme was proposed in \cite{PSO_UAVwsn} to position the UAVs to maximize coverage and sensor data acquisition while maintaining BS connectivity, where a topographical map is known beforehand.
A clustering approach proposed in \cite{c21} uses dual pheromones for target tracking and area coverage, while cluster head (CH) nodes maintain stable network connectivity. 
Relay UAVs may be deployed as a separate layer to provide continuous BS connectivity in a hierarchical network \cite{bs24, Steiner4}. However, the selection and placement of relay nodes add additional complexity and require network topology information.
Moreover, CH or relay UAVs in hierarchical networks can become bottleneck nodes, leading to delays and congestion, and can be vulnerable to attacks \cite{heir_net_challenges2}. 
Our proposed decentralized scheme avoids these issues.

Next, we review reinforcement learning (RL) based schemes for UAV mobility. The highly dynamic and complex nature of UAV networks means that a UAV's actions can have long-term consequences on the system's evolution, making reinforcement learning an appealing framework \cite{survey_FANET_RL}. However, using multi-agent deep reinforcement learning (DRL) algorithms in UAV networks presents several challenges, such as high dimensional states and observations with increasing node density, which make it difficult to find optimal policies \cite{multi-agent1}.  Moreover, communication range constraints make the environment only partially observed to each UAV.
Yue et al. \cite{RL-coverage} use RL for multi-UAV search of targets but do not consider the connectivity. 
In \cite{DRL-CommCoverage}, a group of rotary-wing UAVs act as aerial BS to provide communication coverage to ground users. A distributed UAV control is designed to maximize communication coverage and fairness and minimize UAV energy consumption via DRL. However, the connectivity and coverage trade-off is not evaluated. 
In \cite{bsDRL1}, DRL is used for UAV path planning with connectivity constraints. Here, the UAV learns to fly to its destination in the shortest time while maintaining reliable communication to the ground BS in a cellular network.

We use DRL to optimize the trade-off between area coverage and connectivity (node degree as well as BS connectivity) using only local pheromone and connectivity information. In our scheme, we consider the multi-agent RL problem as a single-agent RL scenario, using centralized training with a distributed evaluation approach.

A few UAV mobility schemes consider the impact of node failures due to the energy consumption \cite{bs22, DRL-CommCoverage}. These are more important for the rotary-wing UAVs (which have a higher power consumption and lower endurance \cite{energy-fixedvsrotary}). Simulations show that our proposed schemes are robust and experience only a limited impact when nodes fail. Our mobility models form a network with built-in redundancy (strong average node degree and BS connectivity). Moreover, UAVs in our network design are not used as CH, gateway or relay nodes and hence the failure of a few nodes does not significantly impact network performance.

\subsection{ConCov Mobility Model}\label{Overview_ConCov}
Given the disadvantages of schemes that use centralized or hierarchical architectures or dedicated relay nodes, we focus on decentralized, non-hierarchical, and autonomous UAV network architectures. There are relatively few such distributed and connectivity-aware UAV mobility models in the literature; we compare our proposed schemes against the ConCov model \cite{bsConCov}, which uses a modified flocking behavior that can trade off between coverage and BS connectivity.
 
Separate models for area coverage (for target detection) and BS connectivity (for notification of targets to BS) were studied in \cite{bsCon}. Later, in \cite{bsConCov}, the author presented the ConCov model which is a distributed, hybrid mobility scheme that focuses on fast area coverage and BS connectivity. The ConCov model utilizes artificial potential fields to minimize the overlap in the area sensed by neighboring UAVs. It also uses local neighbor locations and routing information to tune its performance between area coverage and BS connectivity. 
The heading of a UAV is given by,
\begin{equation}\label{concov_eq}
    \vec{R_v} = \omega \frac{ \vec{R_v^{cov}} }{ |\vec{R_v^{cov}}| } + (1-\omega) \frac{ \vec{R_v^{con}} }{ |\vec{R_v^{con}}| } 
\end{equation}
where $\vec{R_v^{cov}}$ is a ``repellent'' force to reduce the coverage overlap, $\vec{R_v^{con}}$ is an ``attraction'' force to prevent disconnections from the BS, and $\omega$ is a weight to balance the two forces. Each UAV uses its current and next positions, the headings of neighboring UAVs, the resulting network graph, and the route to BS to calculate the two forces.

The repellent force, $\vec{R_v^{cov}}$, is 
defined by a combination of the force vector $\vec{F_{ii}}$ along the current heading direction $\theta_i^c$, and the sum of the repel force vectors $\vec{F_{ij}}$ from its neighbors $j\in N_i^s$ along $\theta_{ij}$. The $\hat{\theta}$ notation is used to represent a unit vector with direction $\theta$.
\begin{equation}\label{eq_Rcov}
    \vec{R_v^{cov}} = \vec{F_{ii}}  +  \sum_{j\in N_i^s} \vec{F_{ij}} 
\end{equation}
\begin{equation}
    \textrm{where,}\quad \vec{F_{ii}}=\frac{1}{r} \hat{\theta_i^c} \quad \textrm{and} \quad \vec{F_{ij}}=\frac{1}{d_{ij}} \hat{\theta_{ij}}
\end{equation}
while $r$ is the UAV's sensing range and $d_{ij}$ is the distance between UAVs $i$ and $j$.

The attraction force, $\vec{R_v^{con}}$, is applied to a UAV $i$ to maintain BS connectivity \cite{bsCon}. If UAV $i$ remains connected to BS after the sensing period $ts$, $\vec{R_v^{con}} = \hat{\theta_i^c}$. If UAV $i$ becomes disconnected from the BS after $ts$, then
\begin{equation}\label{eq_Rcon}
    \vec{R_v^{con}} = \vec{J_{ii}}  +  \vec{J_{ik}}, 
\end{equation}
\begin{equation}
    \textrm{where}\quad \vec{J_{ii}}= \hat{\theta_i^c} \quad \textrm{and} \quad \vec{J_{ik}}= \hat{\theta_{ik}}.
\end{equation}
Here, $\vec{J_{ii}}$ is the vector along the current direction $\theta_i^c$ and $\vec{J_{ik}}$ is the vector from UAV $i$ to its 1-hop neighbor $k$ that has a route to the BS.
The ConCov algorithm is described in Pseudocode \ref{pseudo:concov}. 
\begin{algorithm}[h]
	\SetAlgorithmName{Pseudocode}{}
	\LinesNumbered 
	\setcounter{AlgoLine}{0}
	\begin{small}
        At each UAV $i$ with current direction $\theta_i^c$;\\
		\uIf{time interval of $ts$ has elapsed since last direction change}{
            // Update UAV's direction\\
            Compute repulsive force, $\vec{R_v^{cov}}$ from \eqref{eq_Rcov};\\
            Compute attractive force, $\vec{R_v^{con}}$;\\
            \uIf{there exists a route to BS from UAV $i$ after a time interval of $ts$}{
                $\vec{R_v^{con}} = \hat{\theta_i^c}$;
			    }
			\Else{//No route to BS exists\\
                $\vec{R_v^{con}} = \vec{J_{ii}}  +  \vec{J_{ik}}$ from \eqref{eq_Rcon}; 
			    }
            Compute resultant force $\vec{R_v}$ from \eqref{concov_eq};\\
            // New UAV direction\\
            $\theta_i^c = \angle{\vec{R_v}}$
			}
		\Else{
            Maintain current direction $\theta_i^c$;
            }
  
	\end{small}
	\caption{ConCov}
	\label{pseudo:concov}
\end{algorithm}

\section{Overview of Pheromone Mobility Model}\label{Overview_Pheromone_Mobility_section}
The pheromone mobility model uses repel digital pheromones to promote exploration and fast coverage of an area with no prior information \cite{rajan-b17, CAP_ref, b2}. A digital pheromone has characteristics modeled after natural pheromones, such as deposition, evaporation and diffusion,
with values stored in a digital pheromone map.
We assume the UAVs move in a two-dimensional space (i.e., at constant altitude) to search a given area;
although the UAVs' positions and trajectories are continuous-valued, to represent the pheromone map we 
divide the area into a grid of $C\textsuperscript{2}$ cells, each identified by its $(x,y)$  coordinates.%
\footnote{For convenience, we use the same discrete grid to track our area coverage (in our experimental evaluations), and to define the waypoints used in our trajectory selection policies.}
Each UAV moves towards the cells with minimum repel pheromone value and deposits a repel pheromone of magnitude ‘1’ in the cells scanned along its trajectory. After a UAV deposits pheromone in a cell $(x,y)$, it is progressively diffused to the surrounding cells at a constant diffusion rate $\psi\in[0,1]$. This encourages UAVs to disperse towards unvisited cells. The pheromone value of each cell also evaporates, decreasing its intensity over time by a constant rate $\lambda \in[0,1]$. This evaporation allows the UAVs to revisit already scanned cells after some time gap,
for example if environment or target locations change over time \cite{CAP_ref}.

Mathematically the pheromone value $p_{(x,y)}$ in a cell $(x,y)$ at time $t$ is described as \cite{b2, CAP_ref},
\begin{multline}
p_{(x,y)}(t)=(1-\lambda)  \cdot [ (1-\psi) \cdot p_{(x,y)}(t-1)+\\
\partial p_{(x,y)}(t-1,t) + \partial d_{(x,y)}(t-1,t) ]\label{pher_dynamic}
\end{multline}
where $(1-\psi) \cdot p_{(x,y)}(t-1)$ is the pheromone value remaining in cell $(x,y)$ after diffusion to the surrounding cells, $\partial p_{(x,y)}(t-1,t)$ is the new pheromone value deposited in the update interval $(t-1,t)$, and $\partial d_{(x,y)}(t-1,t)$ is the additional pheromone diffused to the current cell from its eight surrounding cells in the update interval $(t-1,t)$, which is described as \cite{b2, CAP_ref},
\begin{equation}
\partial d_{(x,y)}(t-1,t)=\frac{\psi}{8} \cdot \sum_{a=-1}^1 \sum_{b=-1}^1 p_{(x+a,y+b)}(t-1) \label{pher_diffusion}
\end{equation}

In a decentralized UAV network, the UAVs exchange their digital pheromone maps with their 1-hop neighbors by using the periodic `hello messages' \cite{CAP_ref}.

\section{Node Degree and BS Connectivity Aware Pheromone Mobility Model (BS-CAP)} \label{proposed_BSCAP}
Pheromone mobility models achieve fast area coverage by pushing the UAVs away from each other and recently visited cells. However, this can lead to a weak node degree and BS connectivity due to the limited transmission range of the UAVs, especially when the UAV density is low. Maintaining strong node degree is required in a decentralized 
network to ensure robust network communication and coordination among UAVs. Similarly, strong BS connectivity enables robust transmission of information from UAVs to the BS.

In this section, we describe our proposed BS-CAP mobility model for decentralized, autonomous UAV networks to balance strong average node degree and BS connectivity with fast area coverage. 
Our model uses a weighted combination of the repel pheromone value (see Section \ref{nxtwaypnt} and \ref{lookaheadPher}) and node degree (see Section \ref{dstwtnodedeg}) at the cells, along with `route availability to BS' information, to determine a UAV's trajectory (see Section \ref{nxtwaypntSelection}).
Since it is not practical to know the complete topology of such a network in real-time, the proposed scheme is fully distributed and relies only on neighborhood information (see Section \ref{Hellomessages}).

\subsection{Next-Waypoint}\label{nxtwaypnt}
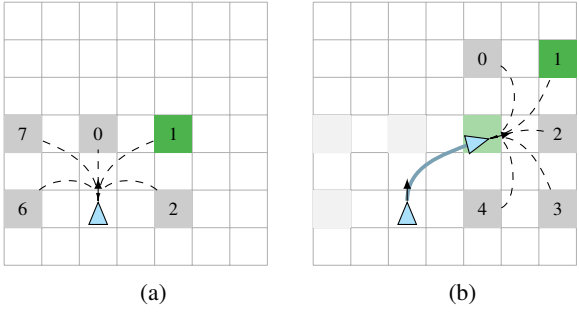
\begin{figure}[t]
\centering
\begin{subfigure}{0.45\linewidth}
\begin{tikzpicture}[scale=0.5]
  \draw[step=1, darkgray!40, thin] (0,0) grid (7,7);
  \fill [gray!40] (0,3) rectangle (1,4); \node (A7) at (.5,3.5) {\scriptsize{7}};
  \fill [gray!40] (0,1) rectangle (1,2); \node (A6) at (.5,1.5) {\scriptsize{6}};
  \fill [gray!40] (2,3) rectangle (3,4); \node (A0) at (2.5,3.5) {\scriptsize{0}};
  \fill [green!40!gray] (4,3) rectangle (5,4); \node (A1) at (4.5,3.5) {\scriptsize{1}};
  \fill [gray!40] (4,1) rectangle (5,2); \node (A2) at (4.5,1.5) {\scriptsize{2}};
  \node[isosceles triangle,draw,inner sep=0,rotate=90,fill=cyan!30,minimum size=.7em] (T) at (2.5,1.25){};
  \draw [->,>=latex,color=black] (T.apex) -- ++(0,.6cm);
  \draw [dashed] (T.apex) to[out=90,in=50] (A6); 
  \draw [dashed] (T.apex) to[out=90,in=-20] (A7); 
  \draw [dashed] (T.apex) to[out=90,in=-90] (A0); 
  \draw [dashed] (T.apex) to[out=90,in=-160] (A1); 
  \draw [dashed] (T.apex) to[out=90,in=130] (A2); 
\end{tikzpicture}
\caption{}
\label{fig:nextwaypoints1}
\end{subfigure}
\begin{subfigure}{0.45\linewidth}
\begin{tikzpicture}[scale=0.5]
  \draw[step=1, darkgray!40, thin] (0,0) grid (7,7);
  \fill [gray!10] (0,3) rectangle (1,4); 
  \fill [gray!10] (0,1) rectangle (1,2); 
  \fill [gray!10] (2,3) rectangle (3,4); 
  \fill [green!40!gray!50] (4,3) rectangle (5,4); 
  \fill [gray!10] (4,1) rectangle (5,2); 
  \node[isosceles triangle,draw,inner sep=0,rotate=90,fill=cyan!30,minimum size=.7em] (Ta) at (2.5,1.25){};
  \draw [color=cyan!30!gray,line width=.5mm] (T.apex) to[out=90,in=-160] (4.25,3.25);
  \draw [->,>=latex,color=black] (Ta.apex) -- ++(0,.6cm);
  \node[isosceles triangle,draw,inner sep=0,rotate=15,fill=cyan!30,minimum size=.7em] (Tb) at (4.25,3.25){};
  \draw [->,>=latex,color=black] (Tb.apex) -- ++(15:.6cm);
  \fill [gray!40] (6,1) rectangle (7,2); \node (B3) at (6.5,1.5) {\scriptsize{3}};
  \fill [gray!40] (6,3) rectangle (7,4); \node (B2) at (6.5,3.5) {\scriptsize{2}};
  \fill [green!40!gray] (6,5) rectangle (7,6); \node (B1) at (6.5,5.5) {\scriptsize{1}};
  \fill [gray!40] (4,1) rectangle (5,2); \node (B4) at (4.5,1.5) {\scriptsize{4}};
  \fill [gray!40] (4,5) rectangle (5,6); \node (B0) at (4.5,5.5) {\scriptsize{0}};
  \draw [dashed] (Tb.apex) to[out=15,in=-20] (B0);
  \draw [dashed] (Tb.apex) to[out=15,in=-110] (B1);
  \draw [dashed] (Tb.apex) to[out=15,in=170] (B2);
  \draw [dashed] (Tb.apex) to[out=15,in=110] (B3);
  \draw [dashed] (Tb.apex) to[out=30,in=0] (B4);  
\end{tikzpicture}
\caption{}
\label{fig:nextwaypoints2}
\end{subfigure}
\caption{
UAV headings and next-waypoints.  (a) Given its current heading, a UAV selects one (green) of the five forward-facing cells out of its eight possible next-waypoint cells (gray). (b) Arriving at this waypoint, the UAV selects its new next-waypoint; discretizing its current heading, waypoint 2 is the closest to straight, so that its forward-facing options are $\{0,1,2,3,4\}$.
}
\label{fig:fig_nextwaypoints}
\end{figure}

As the UAVs fly over the monitored area, we use the pheromone map grid to determine where each UAV should fly next.
The location on the grid toward which a UAV decides to fly is called the 'next waypoint';
the distance between the current and next-waypoints is a function of the UAV speed. 
We restrict our choice of waypoints based on the flight trajectory constraints of a fixed-wing UAV, giving smooth turn trajectories \cite{CAP_ref}.
Concretely, the current (continuous) heading of each UAV (0 to 360 degrees) is discretized into 8 possible directions,
each corresponding to a possible next-waypoint a distance away.
However, for our trajectory, we only consider up to five ``forward-facing'' next-waypoints: the one along the current (discretized) heading,
and two each to the left and right (corresponding to sharper or more gradual turns).
Fig.~\ref{fig:fig_nextwaypoints} illustrates an example of a UAV's next-waypoint cells: initially, the UAV has a current (discretized) heading of ‘0’, and moves to cell `1’ out of the five possible next-waypoint cells (6, 7, 0, 1, 2). Upon reaching cell `1', the UAV has current (discretized) heading direction of ‘2’; it then moves next-waypoint cell ‘1’ out of the possible options (0, 1, 2, 3, or 4).

\subsection{``Look-Ahead'' Pheromone Value}\label{lookaheadPher}
In a basic pheromone mobility model, each UAV moves towards an unvisited region in the map by selecting the cell with the minimum repel pheromone value. In contrast, our scheme computes a `look-ahead pheromone' value, $P'\in[0,1]$, for each potential next-waypoint cell, equal to a weighted average of the repel pheromone at that cell and its eight 1-hop neighbors. Concretely, for next-waypoint cell $(x,y)$,
\begin{equation}
    P^{'}_{(x,y)}=\frac{1}{12} \cdot \Big(3 \cdot P_{(x,y)} +\sum_{a, b\in\{-1,0,1\}} \!\!\!\!\! P_{(x+a,y+b)} \Big) \label{LA-pher}
\end{equation}
where $P_{(x',y')}\in[0,1]$ is the pheromone value in cell $(x',y')$
\cite{CAP_ref}. 

Moving to cells with minimum $P'$ value, the UAV is more likely to steer toward unvisited regions of the map, increasing the area coverage.

\subsection{Distance-Weighted Node Degree} \label{dstwtnodedeg}
The node degree of a UAV represents the number of its 1-hop neighbors. 
Since UAV pairs that are close to their maximum transmission range are more likely to lose their connection in the
near future, 
we calculate a \textbf{distance-weighted connectivity} $(\gamma_{uv})$ between two UAVs $u$, $v$ as a function of their Euclidean distance $d_{uv}$ and transmission range $T_x$. It is defined as \cite{CAP_ref},
\begin{equation} \label{deg}
\gamma_{uv} = 
\scalebox{.9}{$
\begin{cases} 1 & d_{uv} \leq (0.6\cdot T_x) \\
 2.5(1-\frac{d_{uv}}{T_x}) & (0.6\cdot T_x) < d_{uv} \leq T_x \\
 0 & d_{uv} > T_x
\end{cases}
$}
\end{equation}
The \textbf{distance-weighted node degree} $K_u$ of UAV $u$ is defined as the sum of its $\gamma_{uv}$ over its $\cal N$ 1-hop neighbors \cite{CAP_ref}:
\begin{equation}
K_u = \sum_{v \in {\cal N}} \gamma_{uv}.  \label{dst-deg}
\end{equation}
Intuitively, we set $\gamma_{uv}=1$ when the distance between UAVs is within 60\% of the transmission range, because the probability they will remain connected is high. Beyond 60\%, the value $\gamma_{uv}$ decreases linearly with distance $d_{uv}$, discounting more distant neighbors that are less likely to remain connected in the near future.

\subsection{Distributed Information Exchange using `Hello Messages'} \label{Hellomessages}
To share information in the distributed network, 
a Hello message containing each UAV’s updated local information is propagated to its 1-hop neighbors. 
The pheromone value and connectivity information of a UAV’s neighbors is obtained from the Hello messages and used to select the next-waypoint cell. In our scheme, each UAV exchanges Hello messages with its 1-hop neighbors every 2 seconds. Each Hello message is 24 bytes, and consists of the UAV Id (7 bits), its current location (18 bits with 10 m resolution), next-waypoint cell (12 bits for the cell id, assuming 60 x 60 cells of 100 m x 100 m resolution in a 6 km x 6 km area), and a local pheromone map (6-bit pheromone values in the 5 x 5 cells centered at the UAV’s current cell, for 150 bits total). The Hello packet of a UAV also includes the hop count of the shortest route to BS (4 bits). Thus, every UAV can compute the node degree $K_i$ and availability of a route to BS at the next-waypoint cell $i$.

\subsection{Next-Waypoint Selection}\label{Pi_Ki_alpha_Section} \label{nxtwaypntSelection}
Of the five possible next-waypoint cells, we consider only cells where a UAV can maintain a route to the BS, and select a cell $i$ with maximum score $W_i$, where:
\begin{equation} \label{W_i}
W_i = 
\scalebox{1.0}{$
\begin{cases}
    \alpha_i (1-P'_i) & \mbox{if $\exists$ route to BS}\\
    0 & \mbox{if $\nexists$ route to BS}\\
\end{cases}
$}
\end{equation}
where $P'_i$ is the `look-ahead pheromone' value from \eqref{LA-pher} at next-waypoint cell $i$,  
calculated using the UAV's own pheromone map and pheromone information received from its neighbors' Hello messages.
The value $\alpha_i\in[0,1]$ is the UAV's normalized $K_i$ at cell $i$:
\begin{equation} \label{bs-alpha_function}
\alpha_i = 
\scalebox{1.0}{$
\begin{cases}
\frac{K_i}{\beta} & K_i \leq \beta\\
 1  & \beta < K_i \leq \beta'\\
1/3 & K_i > \beta'
\end{cases}
$}
\end{equation}
We compute $K_i$ from \eqref{dst-deg} using the heading information of the UAV's 1-hop neighbors in their most recent Hello messages. Selecting the values $\beta$, $\beta'$ in \eqref{bs-alpha_function} allows the designer to tune the connectivity and coverage balance in our model. For example, selecting a small $\beta$ (e.g., $\beta$ = $0.5$) results in a model with faster area coverage at the cost of connectivity (node degree). On the other hand, a larger $\beta$ (e.g., $\beta$ = $2.5$) results in better node degree at the cost of a higher coverage time.
Intuitively, UAVs with a $K_i = \beta'$ are considered to have strong node degree; even higher values of $K_i$ are discouraged by decreasing $\alpha$ when $K_i > \beta'$ in \eqref{bs-alpha_function}. 
This is desirable behavior since too-high node degrees correspond to congregations of UAVs, which can 
degrade the area coverage performance at low to moderate node densities.

Each UAV maintains connectivity by selecting the next-waypoint with maximum $W_i$ as per \eqref{W_i}, provided a route exists to the BS. 
If no route from the UAV to the BS exists at any of the 5 possible next-waypoint cells, we select the next-waypoint closest to that UAV's 1-hop neighbor which does have a route to the BS.
The next-waypoint selection process of a UAV is described in Pseudocode \ref{pseudo:cap}.
\begin{algorithm}[]
	\SetAlgorithmName{Pseudocode}{}
	\LinesNumbered 
	\setcounter{AlgoLine}{0}
	\begin{small}
		\uIf{UAV reaches next-waypoint cell}{
			// Deposit repel pheromone\\
			Add repel pheromone value = 1 for the current cell in its digital pheromone map\;
			// Select a new next-waypoint cell ($i$)\\
			Identify the next-waypoint cells (out of 5 possible cells), where the UAV can maintain a route to BS\;
            \uIf{ there exists a route to BS from at least one of the next-waypoint cells}{
            Calculate the `look-ahead pheromone’$(P'_i)$ value for each of the next-waypoint cells using \eqref{LA-pher}\;
			Calculate estimated node degree $K_i$ of the UAV at each of the next-waypoint cells via \eqref{dst-deg} \;
			
                Calculate the $W_i$ value for each of these next-waypoint cells using \eqref{W_i}\;
			    Select cell $i = \arg \max_i W_i$ as the UAV's next-waypoint\;
			    }
			\Else{//No route to BS exists from the next-waypoint cells\\
			    Select a next-waypoint cell $i$ closest to the UAV's 1-hop neighbor which has a route to BS.
			    }
			}
		\Else{// Follow smooth-trajectory towards the selected next-waypoint cell}
	\end{small}
	\caption{UAV Next-Waypoint Cell Selection}
	\label{pseudo:cap}
\end{algorithm}

All values required for the calculation of $P_i'$ and $\alpha_i$ are received from the neighbors through Hello messages.
Computation of $P_i'$ at a next-waypoint involves 10 additions and 2 multiplications. Similarly, the computation of $\alpha_i$ involves a maximum of 11 multiplications and 5 additions, assuming 5 1-hop neighbors. Therefore, the computation of $W_i$ for each next-waypoint cell in \eqref{W_i} requires 16 additions and 14 multiplications.
Since our scheme uses up to 5 possible next-waypoint cells, at most 80 additions and 70 multiplications may be required to select the best next-waypoint.

\section{Deep Q-Learning based BS-CAP (BSCAP-DQN) Mobility Model}\label{DQNsection}
In the BS-CAP model, we use local connectivity information (node degree) to augment the pheromone information and allow trade-offs between coverage and connectivity in our network, while information on the availability of routes to the BS from the next-waypoint cells is used to maintain BS connectivity. In this section, we use the framework of reinforcement learning (RL), in particular, deep Q-Learning (DQN) \cite{dqn0-human, ddpg} to train a policy (BSCAP-DQN) for the UAVs that \emph{explicitly} optimizes a trade-off between coverage, node degree and BS connectivity.
RL \cite{rl} attempts to optimize an agent's policy, represented as a distribution over actions given the current state, in order to maximize a (discounted) cumulative reward to that agent over time.  So-called ``online'' RL training methods use agent's current policy to decide what actions to take, influencing the evolution of the system and subsequent experiences used for training.
\footnote{Although the training method is called ``online RL'', our policy is fully trained in simulation beforehand, and held fixed during the actual UAV deployment.  The short lifetime of the UAVs does not allow for significant exploration of policies during any single deployment. In this sense our DQN policy defines a comparable but explicitly optimized alternative to the BS-CAP policy.}
Typically, the experience is gathered by following an ``exploration policy'' which is related to, but may be more random than, the model's current estimate of the best policy; for example, \emph{$\epsilon$-greedy} policies mostly follow the currently estimated best action, but take a random action with probability $\epsilon$ instead.
``Offline'' training \cite{offlineRL}, in contrast, gathers and stores experience by following some (possibly unknown) policy, which can then be used for training the agent's policy \cite{CAPDQN_ref}.

\begin{figure}[]
\centerline{\includegraphics[width=0.77\linewidth]{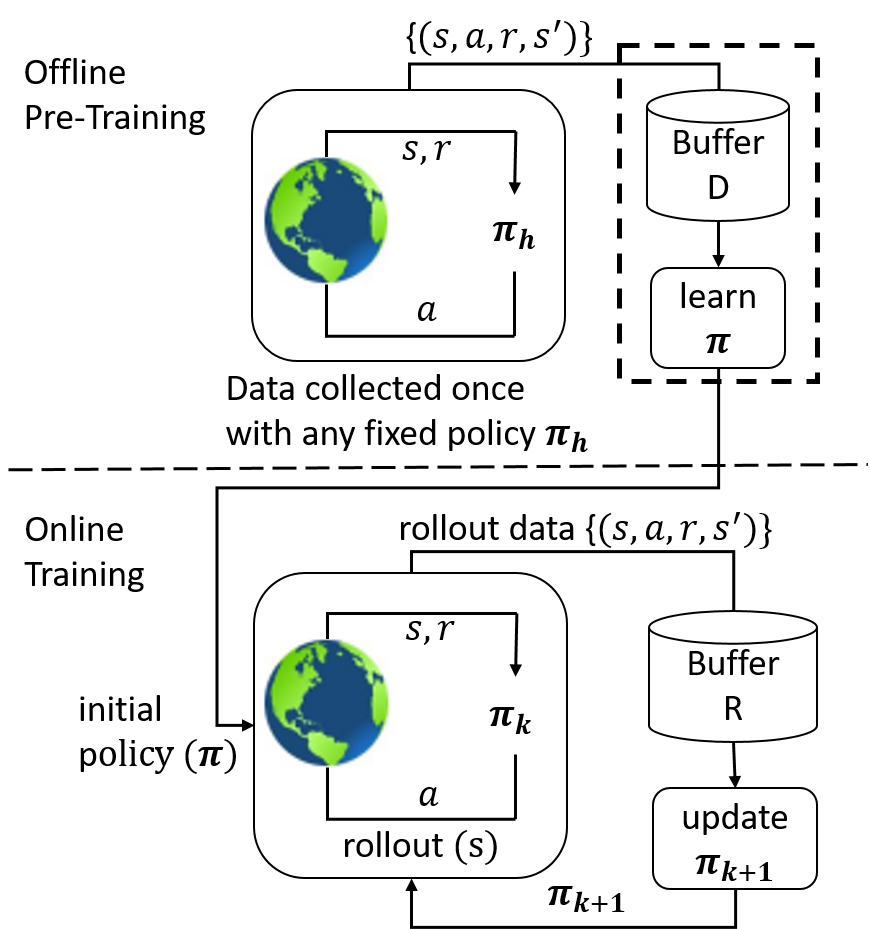}}
\caption{We use offline pre-training, followed by online training, to improve training efficiency while accounting for the effect of policy changes in our multi-UAV setting \cite{CAPDQN_ref}.}
\label{fig_offline-online-RL}
\end{figure}

In practice, 
since the evolution of our system depends on the behavior of multiple UAVs (all following the same policy and using limited state information), it is important to use online training so that changes to the policy can propagate to affect the experience and outcomes observed by other UAVs.  However, starting with a purely random policy leads to slow learning since the UAVs must spend time to learn the rewards associated with exploration.  Thus, to make training more efficient, we first perform pre-training using offline DQN on a database of experience obtained by following an $\epsilon$-greedy exploration variant of our fixed BS-CAP policy from Section~\ref{proposed_BSCAP}.  After this pre-training, we perform online training using DQN with experience replay \cite{dqn0-human} to further refine the agents' policy.  Our implementation uses standard modeling and training libraries from PyTorch \cite{pytorch}. Fig.~\ref{fig_offline-online-RL} illustrates the use of offline pre-training followed by online reinforcement learning \cite{CAPDQN_ref}.

Since our system consists of multiple UAVs, it can be considered an example of multi-agent reinforcement learning \cite{multi-agent1}.  However, the structure of our problem constrains the nature of our agents.  In ``centralized'' multi-agent RL, the agents coordinate through a central decision-maker that incorporates all agents' observations and determines a set of actions jointly.  However, large numbers of agents lead to an exponential explosion in the size of the observation and action spaces.  Moreover, we would like our agents to continue operating even when out of contact with one another, making centralized reasoning unsuitable.

``Concurrent'' multi-agent RL allows each agent to learn its own individual policy.  This allows, for example, agents to learn complementary policies that perform different roles in order to achieve a shared objective.  However, in our setting, we prefer to select a homogeneous policy for the agents (i.e., each UAV follows the same behavior given its local observations), since we would like to apply our policy in settings where there may be more or fewer UAVs than during training.  

Our approach is thus an example of ``parameter-sharing'' multi-agent RL, in which the agents share the parameters of a single policy, which is trained using the experiences of all agents simultaneously (based on each agent's individual observations).  Compared to other multi-agent approaches, this method is highly scalable, and can use a variety of training techniques, including DQN and deep deterministic policy gradient methods \cite{ddpg}.

\subsection{DQN Problem Formulation}
In DQN, we model the state-action value function $q(s,a)$ using a neural network.  The function $q$ captures the long-term value associated with being in a state $s$ and taking action $a$.  Given a current estimate of $q$, our agent can select its next action $a$ to maximize $q(s,a)$ given its current state $s$.

In our setting, each UAV moves from its current waypoint to its selected next waypoint.  Upon reaching the next-waypoint, the UAV must choose a new next-waypoint from a set of five possibilities based on the ``look-ahead'' pheromone ($P'$), distance-weighted node degree ($K$), route availability to BS at those next-waypoints, normalized distances from the next-waypoint cells to the UAV's 1-hop neighbor that has the shortest route to BS, the normalized distance from the UAV to BS, and the node degree of BS.
Thus our \textit{input state} $s$ consists of 22 real values. 
Note that the $P'$ and $K$ help to learn a policy with faster map coverage and better node degree, while route availability to BS at next-waypoint cells, normalized distances from the next-waypoint cells to the 
UAV's 1-hop neighbor with a route to BS, the normalized distance from the UAV to BS, and the node degree of the BS help in learning to maintain strong BS connectivity.

Our \textit{action space} consists of five possible actions corresponding to the five forward-facing directions (next-waypoint cells) with respect to the current UAV heading. 

Our RL setting also requires us to define a set of rewards $r$, which are obtained when the agent takes action $a$ from state $s$.  
We provide three sources of reward: one for area coverage ($r_c$), one for node degree ($r_k$) and one for BS connectivity ($r_b$). The total reward ($r$) is defined as,
\begin{equation} \label{r_function}
    r = m\cdot r_c + r_k + n\cdot r_b
\end{equation}
Here, the coverage reward, $r_c$, is defined as the difference between the number of new cells scanned and the number of already-scanned cells along the path taken by the UAV between its two waypoint cells.  This favors the paths that improve rapid coverage of the environment assuming that the UAVs scan any cells as they pass overhead.  Our node degree reward, $r_k$, is calculated based on $K_i$ of the UAV when it reaches its next waypoint $i$; specifically,

\begin{equation} \label{rk_function}
{
r_k = \scalebox{1.0}{$
    \begin{cases} 
       -1 & 1 < K_i \leq 2 \\
        0 & 2 < K_i < 3 \\
        -4 & \mbox{otherwise.}
    \end{cases}$}
}
\end{equation}
$K_i \leq 2$ incurs a penalty to avoid very low node degree. A value of $2 < K_i < 3$ gives adequate node degree.  
$K_i \geq 3$ incurs a penalty to avoid very high node degree that causes multiple UAVs to congregate, which can degrade the area coverage performance.  

The BS connectivity reward, $r_b$ is given for maintaining connectivity to the BS:
\begin{equation} \label{rb_function}
r_b = 
\scalebox{1.0}{$
\begin{cases}
    0  & \mbox{if $\exists$ route to BS}\\
    -3 & \mbox{if $\nexists$ route to BS}\\
\end{cases}$}
\end{equation}
The model's preferred balance between area coverage and BS connectivity can be changed by varying $m$ and $n$ in the total reward function in \eqref{r_function}.
Q-learning then optimizes the agent's policy to maximize the expected sum of rewards, discounted by a geometric weighting $\gamma^k$ for rewards $k$ steps in the future.  For our environment, we use discount factor $\gamma = 0.9$.

Ideally, we would train our reinforcement agent to directly optimize the performance objectives used to evaluate
in Section~\ref{sec:perfmetrics}.  However, these evaluation metrics are holistic, global run-based quantities that are difficult to use as reward functions due to their long feedback delays, and generating a sufficiently large number of full system trajectories is computationally prohibitive.  For these reasons, we train our agent using local rewards at each step that provide immediate feedback toward behaviors that benefit our overall objectives, learning a good policy faster with fewer runs.
One potential direction for future research is to develop local reward structures that could more closely align with our desired holistic objectives to improve BSCAP-DQN's performance.

Our DQN neural network consists of two dense, fully connected hidden layers with Leaky ReLU activation functions of size 24 and 16 hidden nodes, respectively.  The input layer takes in the length-22 state vector $s$, and the output layer corresponds to 5 Q-function values, i.e., $Q(s,a)$ for each possible action $a$, as described in \textbf{Table~\ref{NN_table}}.  During development, we experimented with a number of other similar network architectures but concluded that this two-layer network performed sufficiently well and used it in all reported experiments.
\setlength{\tabcolsep}{4pt}
\begin{table}[]
\small
\centering 
\caption{Neural Network Description}
\label{NN_table}
\begin{tabular}{|l|l|l|l|}
\hline
\textbf{Layers} & \textbf{Layer type} & \textbf{Size} & \textbf{Activation function} \\ \hline
Input layer & - & (22) & - \\ \hline
Hidden layer 1 & Fully connected & (24) & Leaky ReLU \\ \hline
Hidden layer 2 & Fully connected & (16) & Leaky ReLU \\ \hline
Output layer & - & (5) & - \\ \hline
\end{tabular}
\end{table}

\subsection{Offline Pre-training Using BS-CAP}
Like in \cite{CAPDQN_ref}, an offline static dataset $D$ for pre-training our DQN is generated using a randomized exploration version of our BS-CAP model policy, $\pi_h$. Since our BS-CAP policy is deterministic, we alter it to take a uniformly-at-random action with some probability $\epsilon=0.1$ to ensure that the learner can see the outcome of actions that would not normally be followed by BS-CAP.  We build a dataset consisting of state ($s$), action ($a$), reward ($r$), and next-state ($s’$) transition sequences from 10,000 episodes under simulation settings as described in Section~\ref{SimultionResults}. Each episode is run for 2000 seconds.

For pre-training, we initialize the DQN  with weights using the default (``kaiming\_uniform'') PyTorch weight initialization. We then train on our offline dataset for $N=50$ epochs (passes through the data), using stochastic gradient descent with minibatch size $M=1024$, which are randomly sampled to avoid temporal correlation. We implement our model in PyTorch and use the SGD optimizer with a decaying learning rate initialized to 0.0001 to optimize the squared error, $\big( Q(s_t,a_t) - y_t \big)^2$, 
where $y_t$ is the Bellman target, $y_t = r_t+\gamma \max_{a'} Q(s'_t,a'))$.
The DQN pre-training process is given in Pseudocode \ref{pseudo:dqn}.
\begin{algorithm}[]
	\SetAlgorithmName{Pseudocode}{}
	\LinesNumbered 
	\setcounter{AlgoLine}{0}
	\begin{small}
	    Initialize Q-network with random weights $\theta$\;
	    Initialize target network $\hat{Q}$ with weights $\hat{\theta}=\theta$\;
	    Load Offline Experience $D = \{(s_t,a_t,r_t,s'_t)\}$\;
	    Initialize minibatch size $M$, total epochs $N$\;
	    t=1,U=3000\;
		\For{epoch = $1 \ldots N$}{
		    Partition $D$ into minibatches $\{B_i\}$ of size $M=1024$ at random
			
			\For{each $B_i$}{ 
			    %
			    t=t+1\;
			    \For{each $(s_t,a_t,r_t,s’_t) \in B_i$}{
			        Calculate targets, $y_t=r_t+ \gamma \max_{a'} Q(s'_t,a';\hat \theta)$\;
			    }
			    Calculate MSE loss on minibatch $B_i$: $L_i=\frac{1}{M}\sum_{t} \big(y_t-Q(s_t,a_t;\theta)\big)^2$\;
			    Take a gradient step on $\theta$ at rate $\alpha$\; 

                \If{$t\mod{U}==0$}{
                    Copy Q-network to target network: $\hat \theta = \theta$\;}
		    }
		Update learning rate scheduler $\alpha$\;
	    }
	\end{small}
	\caption{Offline DQN Pre-Training}
	\label{pseudo:dqn}
\end{algorithm}

\subsection{Online Training}
Like in \cite{CAPDQN_ref}, after pre-training, we simulate additional environmental interactions while UAVs follow an $\epsilon$-greedy exploration version of the current DQN policy. We update a single, universal policy from transition sequences obtained from all agents and stored in a replay memory. Samples from the replay memory are then used to update the DQN's value function $Q$.
Online training ensures that changes to the UAV policy are reflected in the behavior of the other UAVs as well, so that the policy will be optimized to perform well within a network of identical agents.

We train the DQN online for a total of $N'$= 4000 episodes and stop each episode at 2000 seconds. During each episode, the transition sequences ($s, a, r, s’$) of individual UAVs are stored into a replay memory of 10,000 transitions. We update our Q-network after every $B'=30$ UAV steps (i.e., transitions saved to replay memory) by sampling a minibatch of size $M'=512$ at random from the replay buffer. During online training, we use the ADAM optimizer to adaptively select the learning rate (initialized to 0.0001), and estimate our target values $y_t$ using a \emph{target network} $\hat Q$ \cite{dqn0-human}, i.e., $y_t = r_t+\gamma \max_{a'} \hat Q(s'_t,a'))$, where $\hat Q$ is a periodically updated copy of $Q$, updated after every $C=100$ gradient steps.
The online DQN training process is shown in Pseudocode \ref{pseudo:dqn-replay}.
\begin{algorithm}[]
	\SetAlgorithmName{Pseudocode}{}
	\LinesNumbered 
	\setcounter{AlgoLine}{0}
	\begin{small}
	    Initialize $\theta$ using offline trained network weights\;
	    Initialize target network $\hat{Q}$ with weights $\hat{\theta}=\theta$\;
	    Initialize replay memory $R$;  $t=1$\;
	    
		\For{episode = $1 \ldots  N'$}{
			\While{$time \leq 2000s$}{
			    UAV selects action $a_t$ via $\epsilon\text{-greedy} \bigl[Q(s_t,a_t;\theta)\bigr]$\;
			    Execute action $a_t$, and observe reward $r_t$ and next state $s’_t$\;
			    Store $(s_t,a_t,r_t,s’_t)$ in $R$\;
			    $t=t+1$\;
			    
			    \If{$t\mod{B'}==0$}{
			        Sample random minibatch of $M'=512$ transitions $(s_j,a_j,r_j,s’_j)$ from $R$\;
    			    
    			    Calculate target Q values, $y_j=r_j+ \gamma \max_{a'}\hat{Q}(s'_j,a_j';\hat \theta)$\;
    			    
    			    Calculate MSE Loss, $L=\frac{1}{M'}\sum_{j=0}^{M'-1}\big(y_j-Q(s_j,a_j;\theta)\big)^2$\;
    			    Take a gradient step on $\theta$ using ADAM\; 
			        
			        \If{$t\mod{B'\cdot C}==0$}{
			            Copy Q-network to target network: $\hat \theta = \theta$\;
			        }
			    }
			 }
	    }
    \end{small}
	\caption{Online DQN Training }
	\label{pseudo:dqn-replay}
\end{algorithm}

Hyperparameters such as episode value, batch size and gradient steps were selected based on empirical trials on small systems and then used in the larger training process. Although training can be slow, it is accomplished beforehand and does not affect the efficiency of the resulting policy. For the neural network in \textbf{Table~\ref{NN_table}}, evaluating the trained DQN policy requires about 992 multiplications and 907 additions to compute the best next-waypoint among the five possibilities.
Since the next-waypoint is computed only periodically, the DQN adds little energy consumption or device complexity.

\section{Simulation Results and Discussion}\label{SimultionResults}
The performance of our proposed BS-CAP and BSCAP-DQN models are compared against the repel pheromone and ConCov \cite{bsConCov} models, discussed in Sections \ref{Overview_Pheromone_Mobility_section} and \ref{Overview_ConCov}, respectively. The UAV network is simulated in Python3, and Table~\ref{sim_table} shows the simulation parameters.

As in our introduction's scenario, we consider a swarm of 30 to 50 low SWaP fixed-wing UAVs (e.g., \cite{locust, switchblade, lancet, coyote}) for monitoring a 6 km $\times$ 6 km area where fixed communication infrastructure, such as a cellular network, is not available.
The aerial BS is located at the bottom center of the map, and the UAVs are launched from the vicinity of the BS. The UAVs are equipped with GPS and have a transmission range of 1 km (e.g., \cite{locust}). We assume a multihop topology with free space propagation (e.g., \cite{ieee80211ah}).
For simplicity, the UAVs are assumed to be point masses, and their mobility is limited to the X-Y plane flying at a constant altitude (typically from 200m to 1km above the ground \cite{switchblade, lancet, coyote}). 
UAVs perform collision avoidance through trajectory modifications.
To facilitate representing the pheromone map, waypoints, and coverage statistics,
the area is divided into grid cells of 100 m $\times$ 100 m each,
consistent with other literature \cite{CAPDQN_ref, b2, bsConCov, camera_spec1, cacoc2}.
A UAV scans the cell in which it currently resides and deposits a repel pheromone of magnitude $1$. We use pheromone evaporation $\lambda$ and diffusion $\psi$ rates of 0.006 each. Each simulation is run for 3000 s, and performance parameters are averaged over 30 simulation runs to study the coverage and connectivity behavior of the network at different stages.

\begin{table}[h]
	\small
	\centering 
	\caption{Simulation Parameters}
	\label{sim_table}
	\begin{tabular}{l l}
	\toprule
		Parameters & Values  \\
	\midrule
		Simulation Time &3000 s \\
		Map Area 		&6 km $\times$ 6 km \\
		Cell Size & 100 m $\times$ 100 m \\
		Transmission Range 	&1 km	\\
		Number of UAVs		&30, 50\\
		UAV Speed 	&20 m/s, 40 m/s	\\
		Evaporation Rate &0.006\\
		Diffusion Rate &0.006\\
        BS Location &Bottom center of map\\
		Number of Runs  &30 \\
  
	\bottomrule
	\end{tabular}\\
\end{table}

Both our models and ConCov~\cite{bsConCov} contain parameters that control the balance of coverage versus connectivity.  In our experiments, we vary these parameters to produce behaviors with differing emphases.
For ConCov, we select values of $\omega \in (0,1)$ in Eq.~\eqref{concov_eq}, which emphasizes connectivity as $\omega$ decreases; we use a sensing period of 5 s to update the UAV headings.  In our BS-CAP model, we vary $\beta \in [0.5, 2.5]$ in Eq.~\eqref{bs-alpha_function}, where increasing $\beta$ results in a higher node degree; there is no explicit tuning parameter for BS connectivity. In our BSCAP-DQN model, the balance between coverage and connectivity, including the BS connectivity, can be varied by setting $n \in [1, 4]$ in the reward function \eqref{r_function}.  

\subsection{Performance Metrics} \label{sec:perfmetrics}
We measure the behavior and performance of our UAV swarm using a number of summarizing statistics.
The \emph{coverage} properties of the swarm are measured by:
\begin{itemize}
    \item \textit{Coverage ($C_v$)}: The average percentage of cells in the map visited by UAVs at least once within a given duration of time; we prefer a higher $C_v$ in a given duration of time.
    \item \textit{Coverage Time ($Tc$)}: The average time taken to scan 90\% of cells in the map. A lower value of $Tc$ indicates faster area coverage.
    \item \textit{Coverage Fairness ($F$)}: Represents how equally all the cells of the map are visited during a given time period, as measured by Jain’s fairness index \cite{Jains_Index},
    \begin{equation} \label{fairness_eq}
        F=\frac{(\sum_i x_i)^2}{n\sum_i x_i^2}
    \end{equation}
    where $x_i$ is the number of scans of cell $i$, and $n$ is the total number of cells in the map. A higher value of $F$ is desired. 
\end{itemize}
The \emph{connectivity} properties are measured by:
\begin{itemize}
    \item \textit{Number of Connected Components ($NCC$)}: Average number of disjoint components in the UAV network (with no path between them), sampled every 10 s. NCC measures how disconnected the network is;  its optimal (minimal) value is 1.
    \item \textit{Average Node Degree ($AND$)}: The node degree of a node $u$ ($N_D(u)$) is its number of links (or 1-hop neighbors). $AND$ is the average node degree of all $V$ nodes in the network, computed by sampling the network every 10 s.
    \begin{equation} \label{AND_eq}
        AND =\frac{\sum_u N_D(u)}{V}
    \end{equation}
    
    \item \textit{Percentage of Time Connected to BS ($T_{bs}$)}: Average percentage of time a UAV is connected, directly or indirectly (through a multihop path), to the BS throughout the simulation.  
    \item \textit{Giant Component ($G$)}: Average size of the connected subgraph (component) with the largest number of nodes in the network; larger $G$ indicates that fewer UAVs are isolated.
\end{itemize}

\subsection{Results and Discussion}
%
\begin{figure*}[]
\centering
\begin{subfigure}{0.49\textwidth}
\begin{tikzpicture}
    \node at (0,0) {{\includegraphics[width=\textwidth]{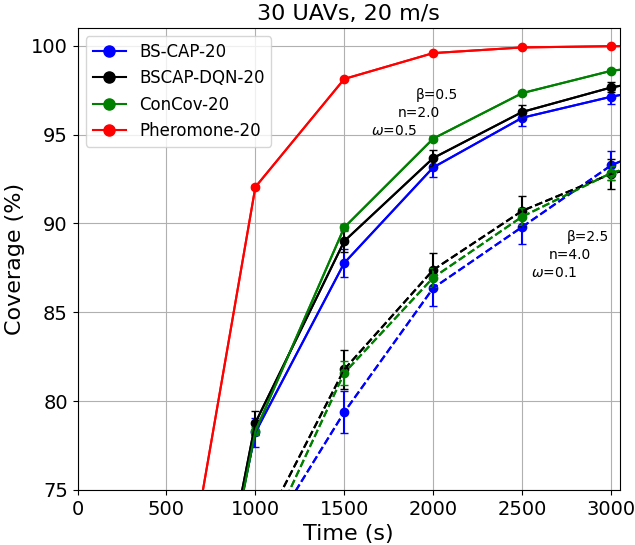}}};
    \draw[->,>=latex,line width=.6mm,brown] (2.7,-2) -- (1.7,-1) node[pos=.1,below] {Better};
\end{tikzpicture}
\caption{}
\label{fig:cov2030}
\end{subfigure}
\begin{subfigure}{0.49\textwidth}
\begin{tikzpicture}
    \node at (0,0) {{\includegraphics[width=\textwidth]{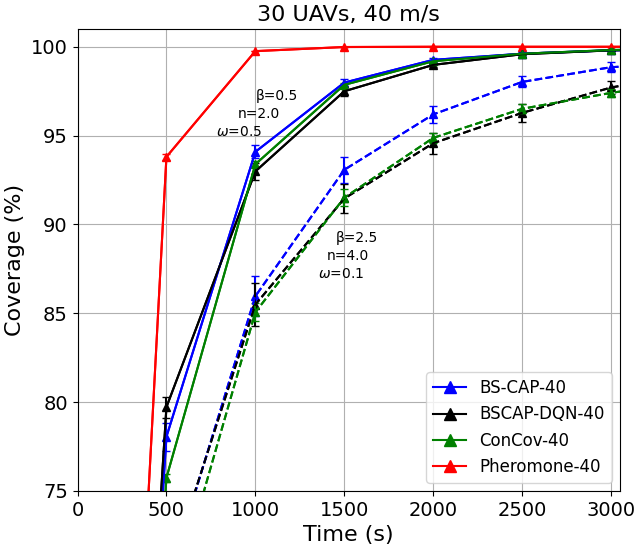} }};
    \draw[->,>=latex,line width=.6mm,brown] (2.2,-.6) -- (1.2,.4) node[pos=.1,below] {Better};
\end{tikzpicture}
\caption{}
\label{fig:cov4030}
\end{subfigure}
\begin{subfigure}{0.49\textwidth}
\begin{tikzpicture}
    \node at (0,0) {{\includegraphics[width=\textwidth]{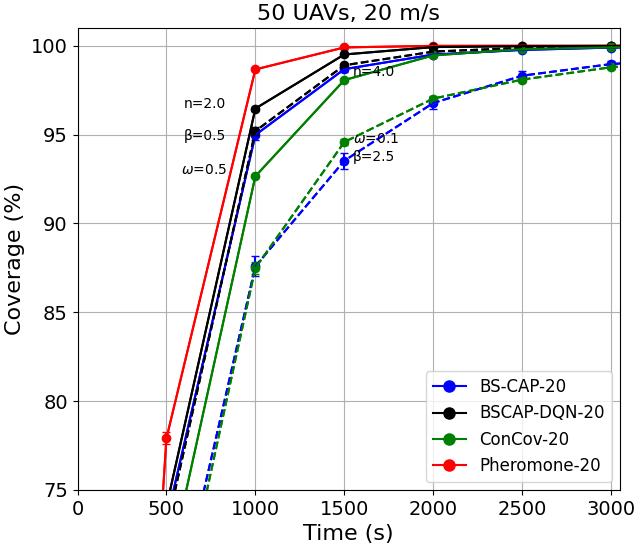} }};
    \draw[->,>=latex,line width=.6mm,brown] (1.7,0) -- (0.7,1) node[pos=.1,below] {Better};
\end{tikzpicture}
\caption{}
\label{fig:cov2050}
\end{subfigure}
\begin{subfigure}{0.49\textwidth}
\begin{tikzpicture}
    \node at (0,0) {{\includegraphics[width=\textwidth]{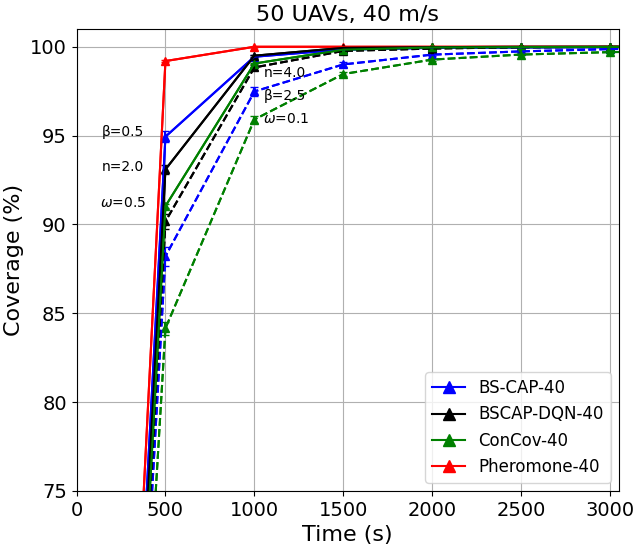} }};
    \draw[->,>=latex,line width=.6mm,brown] (0.6,0.5) -- (-0.4,1.5) node[pos=.1,below] {Better};
\end{tikzpicture}
\caption{}
\label{fig:cov4050}
\end{subfigure}
\caption{Coverage vs. Time plots for 30 and 50 UAVs at 20 m/s and 40 m/s.
While more coverage in less time is preferred, we experiment with several parameter settings to balance coverage with connectivity
(illustrated in subsequent plots);
the parameter values are chosen to produce similar coverage curves among the three tested algorithms (BS-CAP, BSCAP-DQN, and ConCov).
The intermediate parameter settings ($\beta = 1.5$, $n=3$, and $\omega=0.3$) are omitted from these plots for clarity.
}
\label{fig:coverage}
\end{figure*}
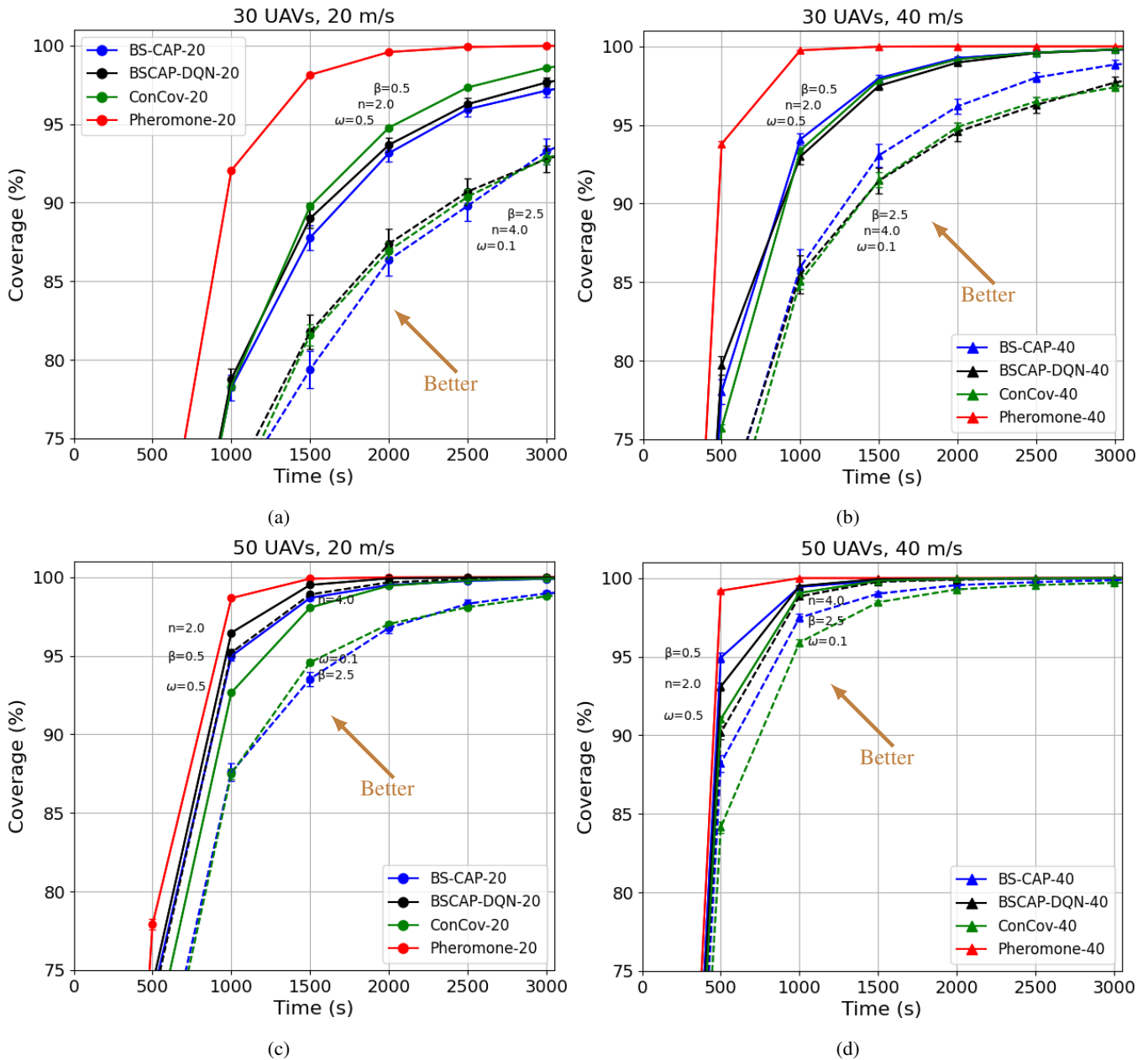

%
\begin{figure*}[]
\centering
\begin{subfigure}{0.49\textwidth}
\begin{tikzpicture}
    \node at (0,0) {{\includegraphics[width=\textwidth]{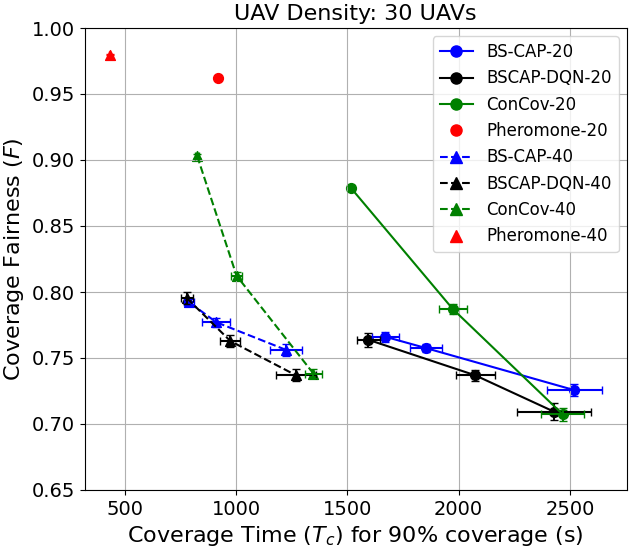}
 }};
    \draw[->,>=latex,line width=.6mm,brown] (-1.5,-2) -- (-2.5,-1) node[pos=.1,below] {Better};
\end{tikzpicture}
\caption{Coverage Fairness}
\label{fig:f30}
\end{subfigure}
\begin{subfigure}{0.49\textwidth}
\begin{tikzpicture}
    \node at (0,0) {{\includegraphics[width=\textwidth]{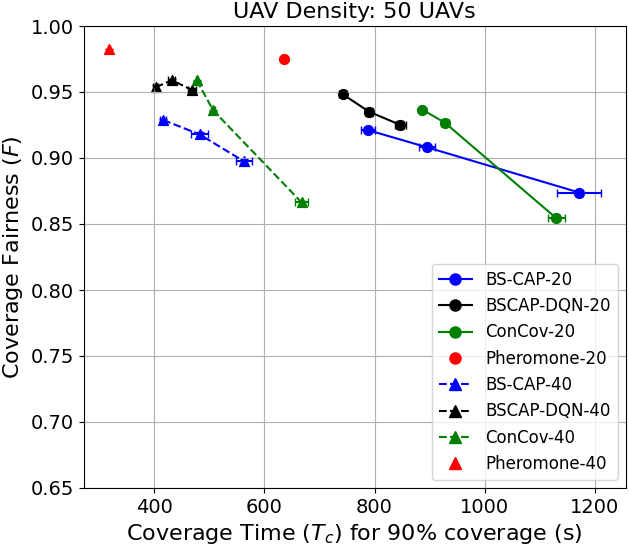}
 }};
    \draw[->,>=latex,line width=.6mm,brown] (0,-0.5) -- (-1,0.5) node[pos=.1,below] {Better};
\end{tikzpicture}
\caption{Coverage Fairness}
\label{fig:f50}
\end{subfigure}
\caption{Coverage Fairness performance plots for 30 and 50 UAVs, at 20 m/s and 40 m/s. (a) For 30 UAVs, ConCov achieves higher $F$ values than BS-CAP and BSCAP-DQN at lower $T_c$ values. The $F$ values of all the models converge when $T_c$ increases. 
(b) For 50 UAVs,  BSCAP-DQN achieves higher $F$ values than BS-CAP and ConCov at lower $T_c$ values.
}
\label{fig:f30and50}
\end{figure*}

\begin{figure*}[]
\centering
\begin{subfigure}{0.45\textwidth}
\par\medskip 
\begin{tikzpicture}
    \node at (0,0) {{\includegraphics[width=\textwidth]{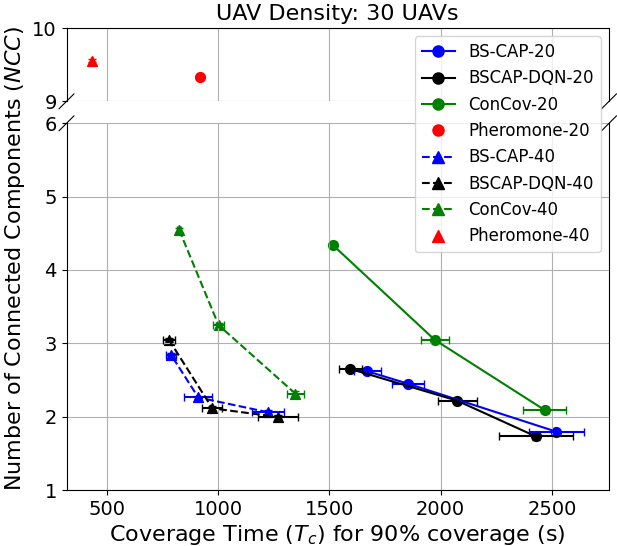} }};
    \draw[->,>=latex,line width=.6mm,brown] (0,1.5) -- (-1,0.5) node[pos=.1,above]
    {Better};
\end{tikzpicture}
\caption{NCC}
\label{fig:ncc30}
\end{subfigure}
\begin{subfigure}{0.45\textwidth}
\begin{tikzpicture}
    \node at (0,0) {{\includegraphics[width=\textwidth]{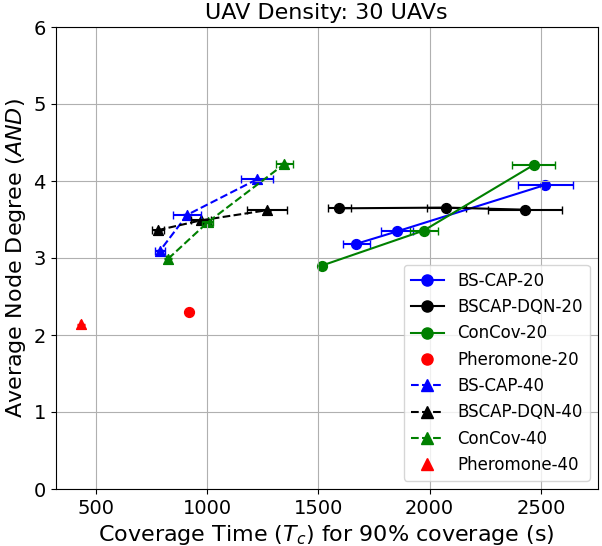} }};
    \draw[->,>=latex,line width=.6mm,brown] (0,-2) -- (-1,-1) node[pos=.1,below] 
    {Better};
\end{tikzpicture}
\caption{AND}
\label{fig:anc30}
\end{subfigure}
\begin{subfigure}{0.45\textwidth}
\par\medskip 
\begin{tikzpicture}
    \node at (0,0) {{\includegraphics[width=\textwidth]{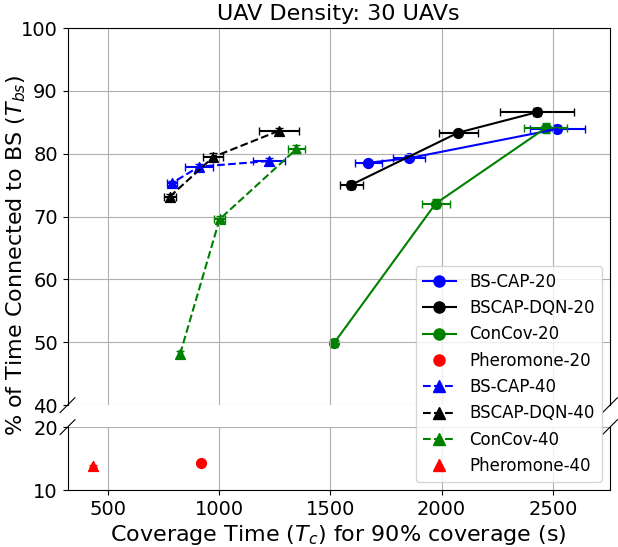} 
    }};
    \draw[->,>=latex,line width=.6mm,brown] (0,-1.5) -- (-1,-.5) node[pos=.1,below] {Better};
\end{tikzpicture}
\caption{ \% Time connected to BS}
\label{fig:tbs30}
\end{subfigure}
\begin{subfigure}{0.45\textwidth}
\begin{tikzpicture}
    \node at (0,0) {{\includegraphics[width=\textwidth]{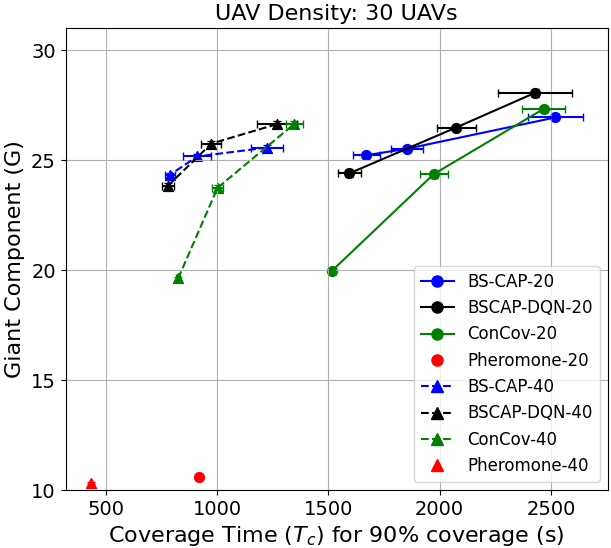} }};
    \draw[->,>=latex,line width=.6mm,brown] (0,-1.5) -- (-1,-.5) node[pos=.1,below] {Better};
\end{tikzpicture}
\caption{Giant Component}
\label{fig:g30}
\end{subfigure}
\caption{Connectivity performance plots for 30 UAVs at 20 m/s and 40 m/s.
(a) At comparable coverage time, BS-CAP and BSCAP-DQN provide a smaller NCC (indicating better connectivity).
The three tested parameter settings for each method are joined by lines, suggesting how intermediate values might fare.
(b) All three methods maintain an average node degree between 3 - 4, providing a sufficient neighborhood.
(c) BS-CAP and BSCAP-DQN provide more time connected to the base station than ConCov at similar coverage times.
(d) Our methods' better connectivity also results in a larger giant component $G$, with fewer isolated UAVs.
}
\label{fig:trade-off-n30}
\end{figure*}
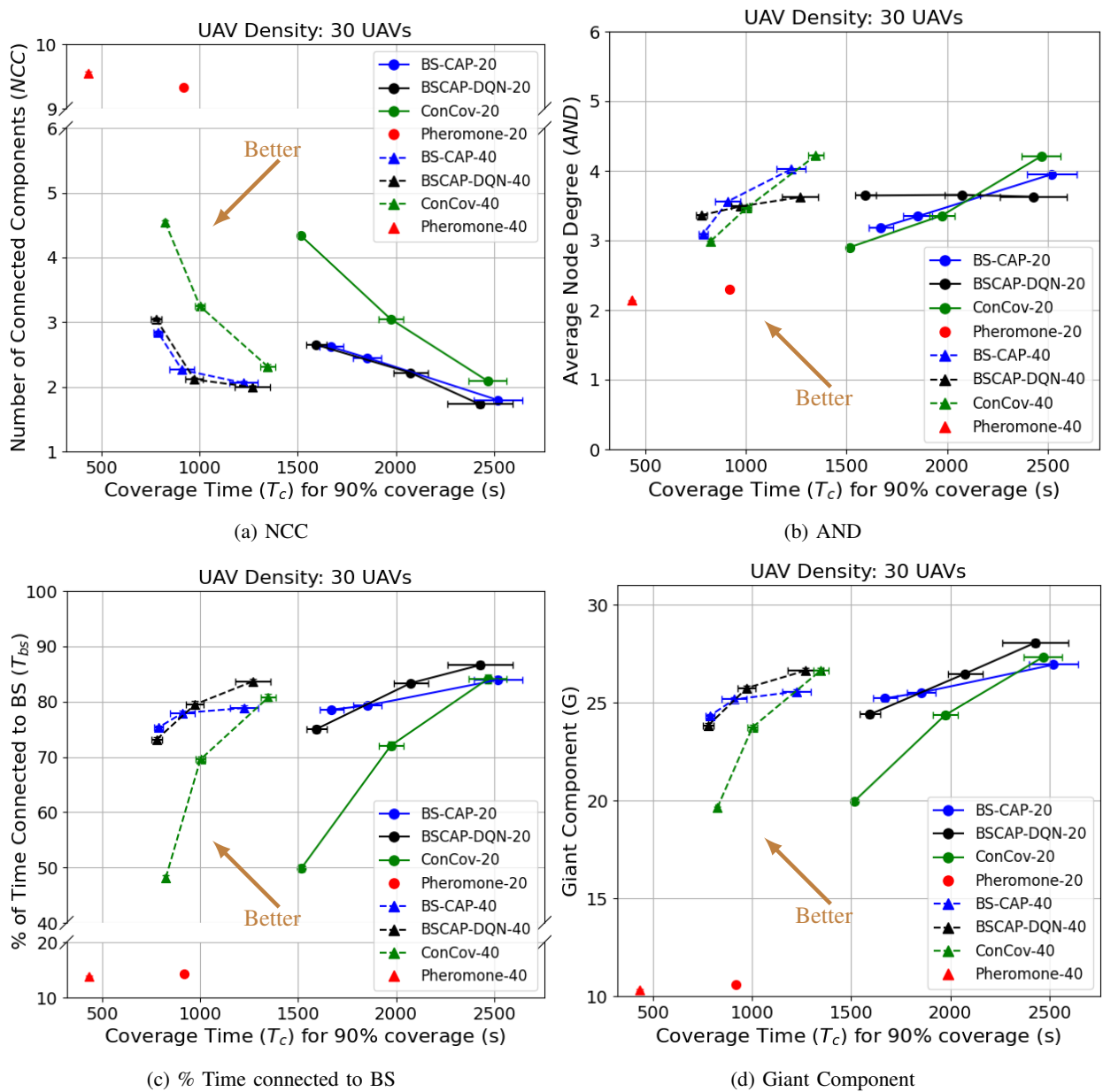
\begin{figure*}[]
\centering
\begin{subfigure}{0.45\textwidth}
\par\medskip 
\begin{tikzpicture}
    \node at (0,0) {{\includegraphics[width=\textwidth]{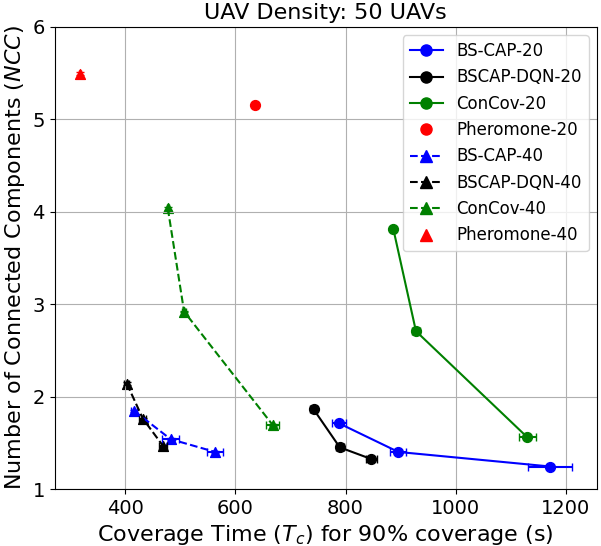}  }};
    \draw[->,>=latex,line width=.6mm,brown] (0,0.5) -- (-1,-0.5) node[pos=.1,above] {Better};
\end{tikzpicture}
\caption{NCC}
\label{fig:ncc50}
\end{subfigure}
\begin{subfigure}{0.45\textwidth}
\begin{tikzpicture}
    \node at (0,0) {{\includegraphics[width=\textwidth]{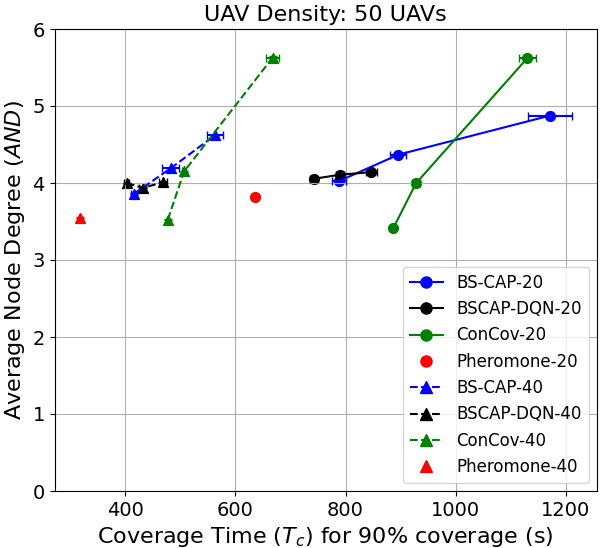} }};
    \draw[->,>=latex,line width=.6mm,brown] (0,-0.5) -- (-1,0.5) node[pos=.1,below] {Better};
\end{tikzpicture}

\caption{AND}
\label{fig:anc50}
\end{subfigure}
\begin{subfigure}{0.45\textwidth}
\par\medskip 
\begin{tikzpicture}
    \node at (0,0) {{\includegraphics[width=\textwidth]{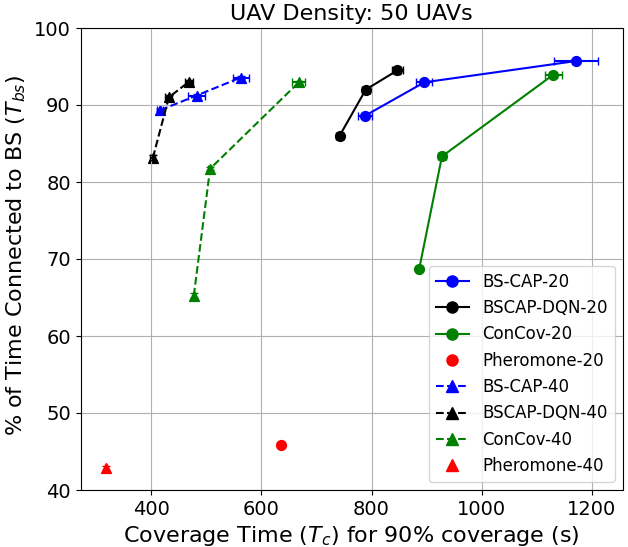} }};
    \draw[->,>=latex,line width=.6mm,brown] (0,-1) -- (-1,0) node[pos=.1,below] {Better};
\end{tikzpicture}

\caption{ \% Time connected to BS}
\label{fig:tbs50}
\end{subfigure}
\begin{subfigure}{0.45\textwidth}
\begin{tikzpicture}
    \node at (0,0) {{\includegraphics[width=\textwidth]{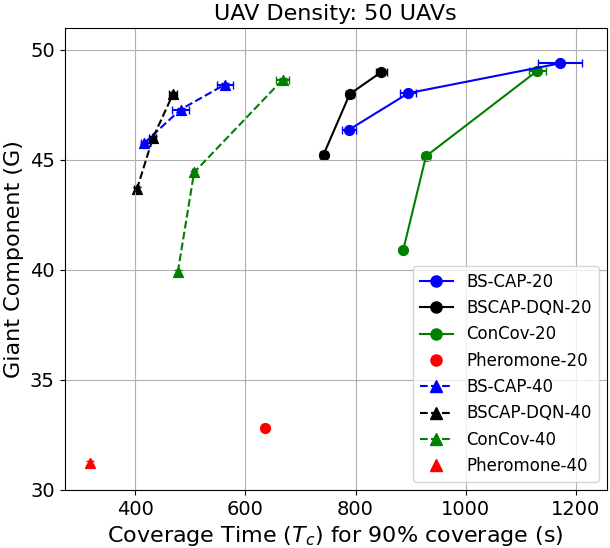} }};
    \draw[->,>=latex,line width=.6mm,brown] (0,-1) -- (-1,0) node[pos=.1,below] {Better};
\end{tikzpicture}

\caption{Giant Component}
\label{fig:g50}
\end{subfigure}
\caption{Connectivity performance plots for 50 UAVs at 20 m/s and 40 m/s.
(a) At comparable coverage time, BS-CAP and BSCAP-DQN provide a smaller NCC (indicating better connectivity).
(b) BS-CAP and BSCAP-DQN maintain an average node degree between 3.9 - 4.9, providing a sufficient neighborhood.
(c) BS-CAP and BSCAP-DQN provide more time connected to the base station than ConCov at similar coverage times.
(d) Our methods' better connectivity also results in a larger giant component $G$, with fewer isolated UAVs.
}
\label{fig:trade-off-n50}
\end{figure*}
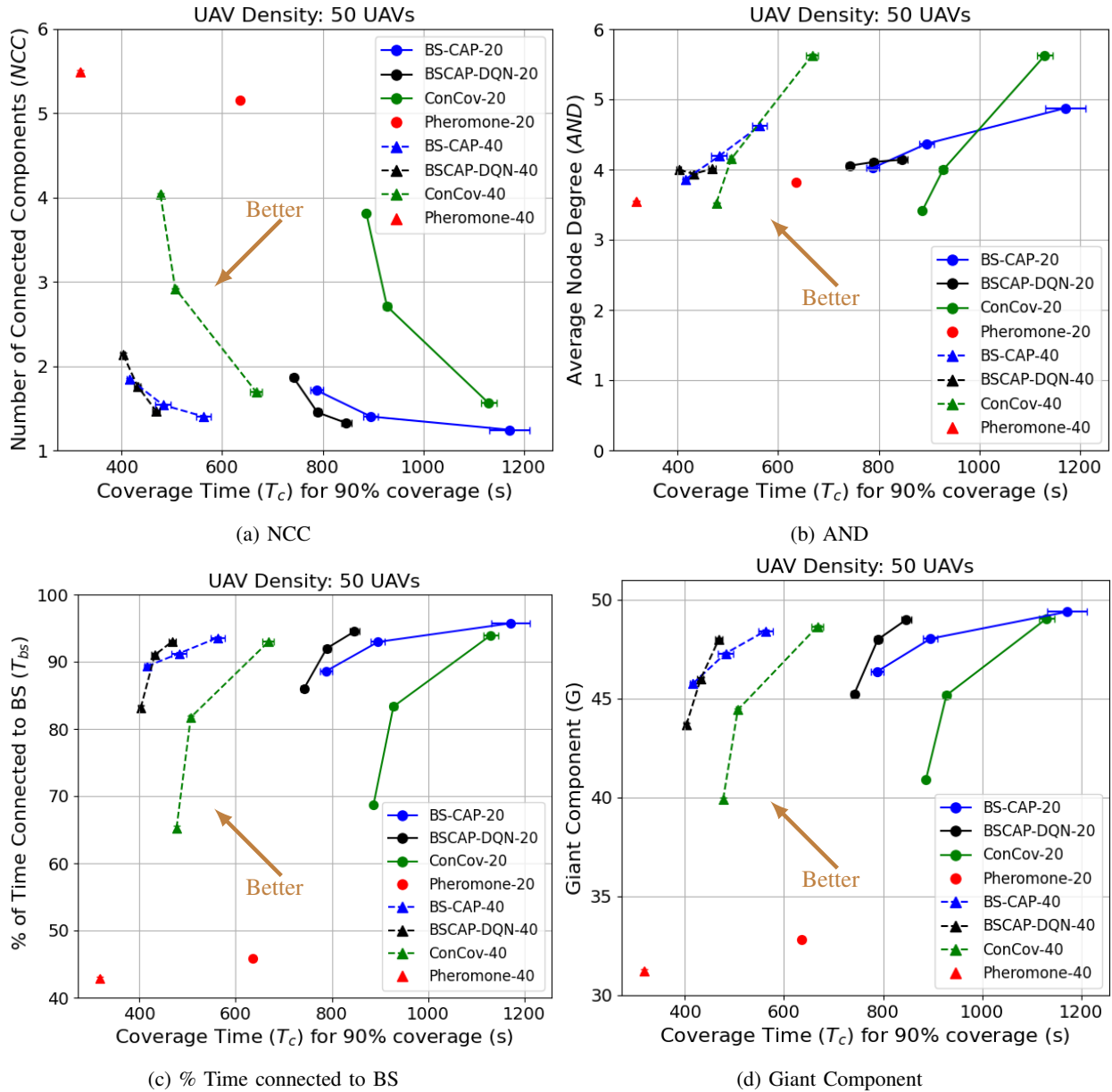

%
\begin{figure*}[]
\centering
\begin{subfigure}{0.49\textwidth}
\begin{tikzpicture}
    \node at (0,0) {{\includegraphics[width=\textwidth]{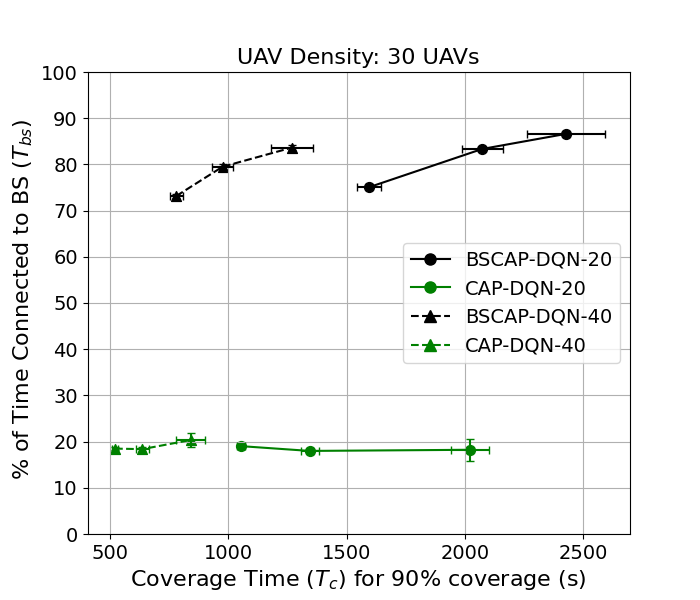}
 }};
    \draw[->,>=latex,line width=.6mm,brown] ((-.5,0) -- (-1.5,1) node[pos=.1,below] {Better};
\end{tikzpicture}

\caption{}
\label{fig:tbs30-cap-dqn-vs-bcap-dqn}
\end{subfigure}
\begin{subfigure}{0.49\textwidth}
\begin{tikzpicture}
    \node at (0,0) {{\includegraphics[width=\textwidth]{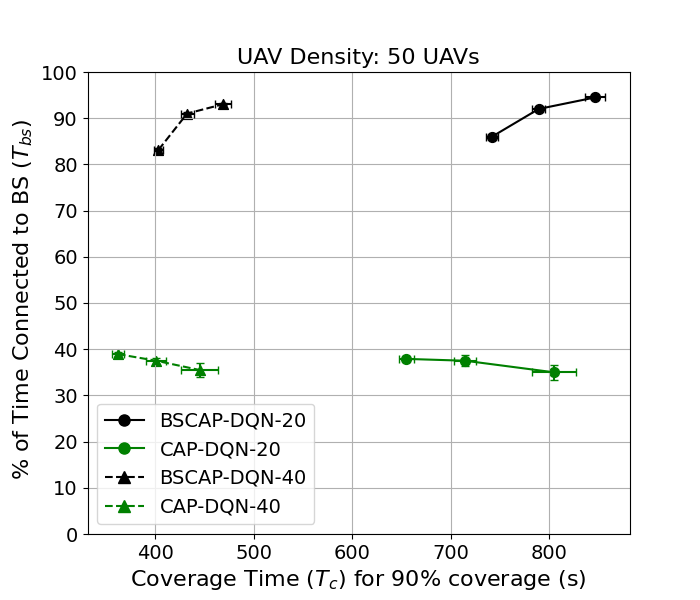} }};
    \draw[->,>=latex,line width=.6mm,brown] ((-.5,0) -- (-1.5,1) node[pos=.1,below] {Better};
\end{tikzpicture}

\caption{}
\label{fig:tbs50-cap-dqn-vs-bcap-dqn}
\end{subfigure}
\caption{$T_{bs}$ BSCAP-DQN vs. CAP-DQN performance plots for 30 and 50 UAVs, at 20 m/s and 40 m/s.
BSCAP-DQN achieves higher $T_{bs}$ than the CAP-DQN since CAP-DQN does not consider the BS connectivity constraints.}
\label{fig:trade-off-cap-dqn-vs-bcap-dqn}
\end{figure*}

We evaluate 
the coverage ($T_c$, $F$) and connectivity ($NCC$, $AND$, and $T_{bs}$) statistics of three connectivity-aware models:
BS-CAP, BSCAP-DQN, and ConCov, as well as 
a basic pheromone model, under four different conditions: densities of 30 or 50 UAVs, at speeds of either 20 m/s or 40 m/s.  
In general, better connectivity comes at the cost of slower coverage; by varying parameters in each scheme, we can trade off between these quantities.
We select three parameter settings for each connectivity-aware method: 
in BS-CAP, we take $\beta \in \{ 0.5, ~1.5, ~2.5\}$ and $\beta'=3$;
in BSCAP-DQN, we take $m=3$ and $n \in \{ 2, ~3, ~4\}$;
and in ConCov, we take $\omega \in \{ 0.5, ~0.3, ~0.1\}$.
Figures \ref{fig:coverage}-\ref{fig:trade-off-cap-dqn-vs-bcap-dqn} show performance plots for each method, 
on average over the 30 runs, with error bars representing the standard error of the averages.
Suffix ``-20'' (e.g., BS-CAP-20, ConCov-20) indicate results at 20 m/s, while 
``-40'' indicates UAVs at 40 m/s speeds.\\

\subsubsection{Coverage Performance}\label{cover_sec}
The \textbf{Coverage vs.~Time plots} in Figure \ref{fig:coverage} represent the total map area covered in a given time.
We illustrate each algorithm with its largest and smallest parameter settings, omitting the middle values to reduce clutter.
As expected, increasing $\beta$ or $n$, or decreasing $\omega$ in their respective models increases the emphasis on connectivity over coverage performance. Increasing node density and/or speed leads to a faster coverage of a given area for all models. By design, at our parameter settings we have relatively similar coverage profiles among the three methods; the basic pheromone model gives the fastest coverage (with no consideration for connectivity). 
Their comparable coverage profiles allow us to see clear distinctions between the methods when comparing their performance under the other metrics,
illustrated in Figures~\ref{fig:f30and50}--\ref{fig:trade-off-n50}.

The \textbf{$\boldsymbol{F}$ vs. $\boldsymbol{T_c}$ performance plots} for 30 UAVs and 50 UAVs at both speeds are shown in Figures \ref{fig:f30} and \ref{fig:f50}, respectively. For 30 UAVs, ConCov achieves a higher $F$ value than our proposed models at lower $T_c$ values, but its connectivity performance is worse (see Section \ref{connect_sec} below). The $F$ values of all the models converge when $T_c$ increases. 
With 50 UAVs,  BSCAP-DQN achieves higher $F$ values than BS-CAP and ConCov for lower $T_c$ values, even though we do not have an explicit reward for fairness in our BSCAP-DQN formulation.\\

\subsubsection{Connectivity Performance}\label{connect_sec}
The number of connected components (NCC) measures the ``disconnectedness'' of the network.
In Figures \ref{fig:ncc30} and \ref{fig:ncc50}, we see \textbf{$\boldsymbol{NCC}$ vs. $\boldsymbol{T_c}$ performance plots} for 30 and 50 UAVs, respectively.
For fast area coverage and strong connectivity, we prefer low $NCC$ along with a low coverage time $T_c$.
For both UAV densities and speeds, BS-CAP and BSCAP-DQN achieve better $NCC$ performance than the ConCov model for a given $T_c$ value. Our optimized BSCAP-DQN gives a similar or slightly better (lower) $NCC$ than our fixed BS-CAP heuristic for a given $T_c$ value. 
For example, in Figure \ref{fig:ncc30} we see that at $T_c \approx$750 s for 30 UAVs, BSCAP-DQN-40 and BS-CAP-40 achieve $NCC$ of 2.9, while ConCov-40 has $NCC \approx$ 4.6. Similarly, 50 UAVs at 40 m/s (Figure~\ref{fig:ncc50}) with policies giving $T_c\approx$ 500 s, BSCAP-DQN-40 and BS-CAP-40 achieve $NCC$ of 1.4 and 1.5, respectively, while ConCov-40 has $NCC \approx$ 2.9. 

Average node degree (AND) captures a more local connectivity assessment.
Figures \ref{fig:anc30} and \ref{fig:anc50} show \textbf{$\boldsymbol{AND}$ vs. $\boldsymbol{T_c}$ performance plots} for 30 and 50 UAVs at both speeds.
For both node densities and speeds, the BS-CAP and BSCAP-DQN achieve reasonably high $AND$ $\geq3$,
maintaining a sufficient number of neighbor UAVs for network connectivity.%
\footnote{We consider that a model with $AND \geq 3$ represents reasonably strong node degree. In fact, a higher value of $AND$ may increase the co-channel interference and the probability of packet collisions during communication.  Moreover, maintaining higher connectivity typically degrades coverage, especially at low UAV densities.}
The ability to tune the BSCAP-DQN model by setting reward $r_k$ in \eqref{rk_function}  allows us to maintain an almost constant $AND$ of around 3.5 and 4 for 30 and 50 UAVs, respectively. In comparison, for 30 UAVs (Figure \ref{fig:anc30}), $AND$ values for BS-CAP models vary from 3 to 4, while ConCov models vary from 2.9 to 4.2. Similarly, for 50 UAVs (Figure \ref{fig:anc50}), BS-CAP's $AND$ values vary from 3.9 to 4.9, while ConCov's vary from 3.5 to 5.6.

Perhaps most importantly, we would like to maintain connection to the BS as much as possible.
The \textbf{$\boldsymbol{T_{bs}}$ vs. $\boldsymbol{T_c}$ performance plots} are shown in Figures \ref{fig:tbs30} and \ref{fig:tbs50}. 
We see that BS-CAP and BSCAP-DQN achieve higher BS connection time $T_{bs}$ than ConCov, at smaller $T_c$ values (faster coverage) for both settings of density and speed. For example, at $T_c \approx$ 2000 s for 30 UAVs at 20 m/s, BSCAP-DQN-20 and BS-CAP-20 achieve $T_{bs}$ of above 80\%, compared to 72\% by ConCov-20.  
When we tune the models to achieve higher $T_{bs}$ (stronger BS connectivity), the coverage time $T_c$ increases due to the stricter connectivity constraints. We find that the BSCAP-DQN model outperforms the BS-CAP model at higher $T_{bs}$,
perhaps because the value of $n$ in \eqref{r_function} allows more direct control over the coverage vs. BS connectivity trade-off in BSCAP-DQN,
compared to BS-CAP.

The \textbf{$\boldsymbol{$G$}$ vs. $\boldsymbol{T_c}$ performance plots} for 30 and 50 UAVs are shown in Figures \ref{fig:g30} and \ref{fig:g50}, respectively. At both densities, BS-CAP and BSCAP-DQN achieve a larger giant network component (G) than ConCov for faster coverage times (low $T_c$).
A larger value of $G$ indicates that fewer isolated UAVs are present in the network; its increase is related to our methods' improved NCC and $T_{bs}$ connectivity statistics.
For example, at $T_c \approx$ 800 s for 30 UAVs at 40 m/s (Figure \ref{fig:g30}), BS-CAP and BSCAP-DQN achieve a $G$ value of 24, compared to ConCov's $G$ value of 19.5. The component size $G$ improves in all three models as we increase the emphasis on connectivity. 

\textbf{Comparing $\boldsymbol{T_{bs}}$ vs. $\boldsymbol{T_c}$ Performance of CAP-DQN and BSCAP-DQN}: To study the impact of considering the BS connectivity on coverage performance, we compare the coverage ($T_c$) and BS connectivity ($T_{bs}$) performance of BSCAP-DQN model with 
an earlier version which did not consider BS connectivity, denoted CAP-DQN \cite{CAPDQN_ref}.
We find that BSCAP-DQN achieves around 300\% and 125\% increase in $T_{bs}$ compared to the CAP-DQN policy for 30 and 50 UAVs, respectively; see Figure \ref{fig:trade-off-cap-dqn-vs-bcap-dqn}. Since CAP-DQN has no BS connectivity constraints, it provides faster coverage times ($T_c$) with good node degree, but fails to maintain a strong BS connectivity.\\

\subsubsection{Impact of Node Failures}\label{fail_sec}
Low SWaP UAVs are prone to failure due to mechanical malfunctions or energy consumption. In this section, we study the area coverage and connectivity performance of our mobility models when a fraction of nodes randomly fail during the simulation time of 2000 s. We consider networks at both densities (30 and 50 nodes), at a speed of 20 m/s. Nodes fail progressively with 10\% and 30\% nodes failing by the end of simulation. Other simulation parameters are kept as in Table \ref{sim_table}. We evaluate using parameter values:  $\beta=1.5$ and $\beta'=3$ in BS-CAP,
$m=3$ and $n=3$ in BSCAP-DQN, and $\omega=0.3$ in ConCov.

\begin{figure*}[]
\centering
\begin{subfigure}{0.49\textwidth}
\begin{tikzpicture}
    \node at (0,0) {{\includegraphics[width=\textwidth]{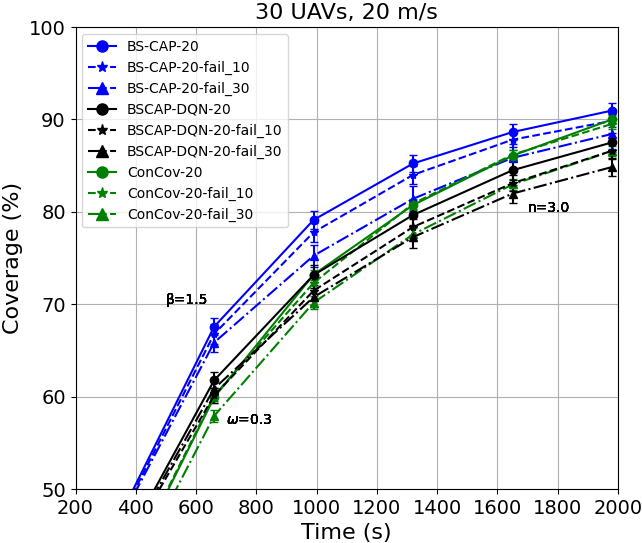}}};
    \draw[->,>=latex,line width=.6mm,brown] (2,-1.5) -- (1,-0.5) node[pos=.1,below] {Better};
\end{tikzpicture}
\caption{}
\label{fig:cov2030-drop}
\end{subfigure}
\begin{subfigure}{0.49\textwidth}
\begin{tikzpicture}
    \node at (0,0) {{\includegraphics[width=\textwidth]{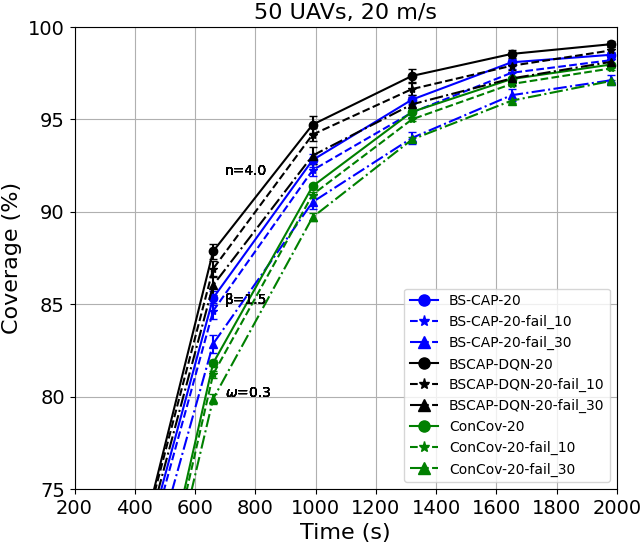} }};
    \draw[->,>=latex,line width=.6mm,brown] (2.2,0.3) -- (1.2,1.3) node[pos=.1,below] {Better};
\end{tikzpicture}
\caption{}
\label{fig:cov2050-drop}
\end{subfigure}

\caption{Coverage vs. Time plots for 30 and 50 UAVs at 20 m/s with 0\%, 10\% and 30\% node failure. UAV failures result in decreased coverage performance.
BS-CAP, BSCAP-DQN and ConCov use parameter settings of $\beta = 1.5$, $n = 3$ and $\omega=0.3$, respectively; for 50 UAVs, BSCAP-DQN-20 uses $n = 4$ in order to show comparable coverage performance. 
Suffix ``-fail\_10'' and ``-fail\_30'' indicate 10\%  and 30\% node failures, respectively.
}
\label{fig:coverage-drop}
\end{figure*}

The \textbf{Coverage vs. Time plots} in Figure \ref{fig:coverage-drop} show the total map area covered in a given time by the three mobility models. We see coverage performance decreases gracefully, in proportion to the number of node failures. For example, the area coverage by BS-CAP decreases from $\approx$80\% to $\approx$78.5\% and $\approx$76\%, when 10\% and 30\% nodes fail, respectively.

In Table \ref{table-drop}, we show the connectivity ($NCC$, $AND$, $T_{bs}$, $G$) and coverage fairness ($F$) performance for 10\% and 30\% node failures. Both $NCC$ and $AND$ are hardly impacted, while $T_{bs}$, $G$ and $F$ decrease gracefully in proportion to the number of node failures. This indicates all three mobility models are relatively robust to node failure.

\begin{table*}[]
\caption{Performance metrics of BS-CAP, BSCAP-DQN and ConCov for different percentages of UAV failure using parameter settings of $\beta = 1.5$, $n = 3$ and $\omega=0.3$, respectively; for 50 UAVs, BSCAP-DQN-20 uses $n = 4$.}

\centering
\setlength{\tabcolsep}{10pt} 
\renewcommand{\arraystretch}{1.2} 
\begin{tabular}{|c|c|c|c|c|c|c|c|}
\Xhline{1.8pt}
\multirow{2}{*}{\textbf{}} &
  \multirow{2}{*}{\textbf{\begin{tabular}[c]{@{}c@{}} \% Node Failures\\ \end{tabular}}} &
  \multicolumn{3}{c|}{\textbf{30 UAVs, 20 m/s}} &
  \multicolumn{3}{c|}{\textbf{50 UAVs, 20 m/s}} \\ \cline{3-8} 
 &
   &
  \textbf{BS-CAP} &
  \textbf{BSCAP-DQN} &
  \textbf{ConCov} &
  \textbf{BS-CAP} &
  \textbf{BSCAP-DQN} &
  \textbf{ConCov} \\ \Xhline{1.8pt}
\multirow{3}{*}{\textbf{NCC}}   & \textbf{0}  & 2.3  & 2.1 & 3 & 1.4 & 1.3 & 2.7 \\ \cline{2-8} 
                                & \textbf{10} & 2.3  & 2.2 & 3 & 1.5 & 1.4 & 2.8 \\ \cline{2-8} 
                                & \textbf{30} & 2.4  & 2.3 & 3.1 & 1.6 & 1.5 & 3 \\ \Xhline{1.8pt}
\multirow{3}{*}{\textbf{AND}}   & \textbf{0}  & 3.5  & 3.8 & 3.4 & 4.4 & 4.2 & 4.1 \\ \cline{2-8} 
                                & \textbf{10} & 3.4  & 3.8 & 3.4 & 4.3 & 4.2 & 4.1 \\ \cline{2-8} 
                                & \textbf{30} & 3.3  & 3.8 & 3.3 & 4.2 & 4.2 & 4 \\ \Xhline{1.8pt}
\multirow{3}{*}{\textbf{Tbs \%}} & \textbf{0}  & 80   & 84 & 72 & 94 & 95 & 84 \\ \cline{2-8} 
                                & \textbf{10} & 76   & 80 & 69 & 91 & 91 & 80 \\ \cline{2-8} 
                                & \textbf{30} & 70   & 73 & 61 & 84 & 84 & 70 \\ \Xhline{1.8pt}
\multirow{3}{*}{\textbf{G}}     & \textbf{0}  & 26   & 27 & 24 & 48 & 49 & 45 \\ \cline{2-8} 
                                & \textbf{10} & 25   & 25 & 23 & 46 & 46 & 43 \\ \cline{2-8} 
                                & \textbf{30} & 22   & 23 & 22 & 42 & 42 & 39 \\ \Xhline{1.8pt}
\multirow{3}{*}{\textbf{F}}     & \textbf{0}  & 0.76 & 0.74 & 0.78 & 0.91 & 0.92 & 0.92 \\ \cline{2-8} 
                                & \textbf{10} & 0.74 & 0.73 & 0.77 & 0.89 & 0.91 & 0.91 \\ \cline{2-8} 
                                & \textbf{30} & 0.72 & 0.71 & 0.74 & 0.86 & 0.88 & 0.88 \\ \Xhline{1.8pt}
\end{tabular}
\label{table-drop}
\end{table*}

In conclusion, at both densities of UAVs and both speeds, the pheromone model provides the best coverage performance ($T_c,~ F$), but has the worst connectivity performance ($NCC,~ AND,~ T_{bs},~ G$) since it does not consider any connectivity information.  Our proposed BSCAP-DQN and BS-CAP models outperform the pheromone and ConCov models by providing better connectivity performance (lower NCC, higher AND, and higher $T_{bs}$).  BSCAP-DQN and BS-CAP provide better connectivity at a similar coverage performance compared to ConCov.  Finally, we find that our RL-based BSCAP-DQN model performs only slightly better than our heuristic BS-CAP model, suggesting that our heuristic weighting performs reasonably well. BSCAP-DQN and BS-CAP are relatively robust to the node failures. 

\section{Conclusion} \label{conclusion_section}
We considered a decentralized, multi-hop UAV network consisting of low SWaP fixed-wing UAVs. 
When monitoring an area autonomously, the area coverage and connectivity requirements of the UAV network exhibit a fundamental trade-off. 
Although fast area coverage is needed to quickly scan the area, strong node degree and BS connectivity are needed to coordinate the UAVs and transmit sensed information to the BS. To facilitate reliable communication among UAVs and to the BS in an autonomous UAV network, we designed a connectivity-aware pheromone (BS-CAP) mobility model. We then developed a deep Q-learning policy based BS-CAP model (BSCAP-DQN). Both BS-CAP and BSCAP-DQN facilitate efficient area coverage while maintaining strong node degree and BS connectivity, and significantly improve over existing schemes.
Our proposed schemes work online, are fully distributed, rely only on neighborhood information, and are robust to node failures, making them practical for real-time coordination in a decentralized UAV network.

Our RL-based model provides slightly better performance than our heuristic BS-CAP model.  This both suggests that the BS-CAP heuristic performs reasonably well, and that to improve significantly further we may need to incorporate more information into the state representation of our RL agent.  For example, we could expand the state to include a history of pheromone and connectivity observations or additional information (obtained in a distributed manner) from the UAV's neighbors, such as their recent trajectories or their own pheromone information.  We leave these as avenues for future research.

\section{ACKNOWLEDGMENT OF SUPPORT AND DISCLAIMER}
This material is based on research sponsored by the Air Force Research Laboratory under agreement No. FA8750-18-1-0023. The views and conclusions contained herein are those of the authors and should not be interpreted as necessarily representing the official policies or endorsements, either expressed or implied, of the Air Force Research Laboratory or the U.S. Government.

\end{document}